\documentclass[useAMS,usenatbib]{mn2e}

\pdfoutput=1

\usepackage{graphicx}
\usepackage{aas_macros}
\usepackage{amsmath}
\usepackage{amssymb}
\usepackage{latexsym}
\usepackage{wasysym}
\usepackage{url}
\usepackage{hyperref}

\title[The Persistent Percolation of Single-Stream Voids]{The Persistent Percolation of Single-Stream Voids}
\author[Falck \& Neyrinck]{B. Falck,$^1$\thanks{E-mail:bridget.falck@port.ac.uk}
	M. C. Neyrinck,$^2$ \\
	$^1$Institute of Cosmology and Gravitation, University of Portsmouth, Dennis Sciama Building, Burnaby Rd, Portsmouth PO1 3FX, UK \\
	$^2$Department of Physics and Astronomy, Johns Hopkins University, 3400 N Charles St, Baltimore, MD 21218, USA}
\date{Draft version \today}

\begin{document}

\label{firstpage}

\newcommand{\figscl}{0.3}
\newcommand{\org}{{\scshape origami}}
\newcommand{\ihmpc}{\,$h$\,Mpc$^{-1}$}
\newcommand{\hmpc}{\,$h^{-1}$\,Mpc}
\newcommand{\hmsun}{\,$h^{-1}\,M_{\astrosun}$}
\newcommand{\reff}{$R_{\rm eff}$}

\maketitle

\begin{abstract}

We study the nature of voids defined as single-stream regions that have not undergone shell-crossing. We use \org\ to determine the cosmic web morphology of each dark matter particle in a suite of cosmological $N$-body simulations, which explicitly calculates whether a particle has crossed paths with others along multiple sets of axes and does not depend on a parameter or smoothing scale. The theoretical picture of voids is that of expanding underdensities with borders defined by shell-crossing. We find instead that locally underdense single-stream regions are not bounded on all sides by multi-stream regions, thus they percolate, filling the simulation volume; we show that the set of multi-stream particles also percolates. This percolation persists to high resolution, where the mass fraction of single-stream voids is low, because the volume fraction remains high; we speculate on the fraction of collapsed mass in the continuum limit of infinite resolution. By introducing a volume threshold parameter to define underdense void ``cores'', we create a catalog of \org\ voids which consist entirely of single-stream particles and measure their percolation properties, volume functions, and average densities.

\end{abstract}


\section{INTRODUCTION}

The large scale structure of the Universe displays a complex, interconnected, hierarchical ``cosmic web'' composed of halos, filaments, walls, and voids~\citep{Klypin1983,Bond1996}. These distinctions reflect the nature of the gravitational collapse of a nearly smooth primordial density field: halos, filaments, and walls are undergoing collapse along three, two, and one dimensions, respectively, while voids are expanding in three dimensions~\citep{Zeldovich1970,Shandarin1984}. The result is that voids dominate the volume of the Universe while containing only a small fraction of the mass, while halos contain most of the mass and occupy only a small fraction of the volume. Voids are not empty, however: there is a hierarchy of smaller voids nested within large voids, as well as small-scale filaments, walls, and halos residing within voids. This hierarchical structure of voids is predicted by theoretical models~\citep[e.g.,][]{Sahni1994,ShethR2004}, measured in numerical simulations~\citep[e.g.,][]{vandeW1993,Gottlober2003,Aragon2013}, and observed in galaxy surveys~\citep[e.g.,][]{Pan2012,Sutter2012Voids,Beygu2013,Alpaslan2014}.

The pioneering work of~\citet{ShethR2004} used the excursion set formalism to develop an analytical model for the distribution and evolution of voids in hierarchical scenarios of structure formation, following the success of this approach in modeling the distribution and masses of halos~\citep{Lacey1993,Sheth1998,Sheth2001}. Excursion set models of voids have been improved and expanded by several authors~\citep{Furlanetto2006,Paranjape2012,Jennings2013,Achitouv2013} and recently applied to modified gravity cosmological models~\citep{Clampitt2013,Lam2014}. 
The excursion set formalism calculates at what scale a random walk in the initial Gaussian random field crosses a threshold density value -- the so-called first crossing distribution. For voids, this threshold value relates to shell-crossing: as the interior of the void expands faster, matter accumulates at the boundary and eventually inner shells pass outer shells. For isolated spherical top-hat underdensity profiles in an Einstein-de Sitter universe, shell-crossing happens at a linearly extrapolated density threshold of $\delta= - 2.81$~\citep{Blumenthal1992,Dubinski1993,ShethR2004}. In this picture, then, voids are single-stream regions, in which the velocity field is single-valued, surrounded by a multi-stream boundary ridge.

The hierarchical structure of voids is described by the \textit{void-in-void} process, whereby the random walk crosses the void shell-crossing density threshold at two different scales; this indicates a sub-void living within a larger void. This is analogous to the \textit{cloud-in-cloud} process of subhalos within larger halos. A second process, unique to voids, is the \textit{void-in-cloud} process, whereby the random walk crosses the halo collapse barrier at a large scale and the void shell-crossing barrier at a smaller scale, resulting in a void that lives within a larger overdensity. This void will eventually become squeezed out of existence as the larger overdensity collapses~\citep{Sutter2014}. The fact that the barriers are constant is a reflection of the spherical expansion (or collapse) assumption. Though~\citet{ShethR2004} (and others) recognized that voids need not be spherical, it is nevertheless taken to be a reasonable assumption because isolated voids become more spherical as they grow~\citep{Icke1984,Bertschinger1985}. However, \citet{Shandarin2006} cast doubts on the validity of the spherical assumption, and recently~\citet{Achitouv2013} showed that spherical evolution is not followed in detail for voids found in the dark matter density field.

Our understanding of the structure and dynamics of voids has greatly increased through the use of cosmological $N$-body simulations. We can measure their profiles~\citep{Ricciardelli2014,Hamaus2014Profile,Nadathur2014}, determine their effect on the CMB~\citep{Cai2014,Hotchkiss2014}, search for new probes of cosmology~\citep{Lavaux2010,Bos2012,Li2012,Hamaus2014Corr}, and investigate the properties of dark matter halos and galaxies within voids~\citep{Rieder2013}. However, there are many different ways to define voids in simulations~\citep[see][and references therein]{Colberg2008}, just as there are many different techniques to find halos~\citep{Knebe2011,Knebe2013}. In contrast to the theoretical expectation that void boundaries are defined by shell-crossing, voids in these methods are identified in the dark matter density field; some also use information in the velocity field, and some use halos as tracers of density instead of dark matter particles. For example, in watershed methods void boundaries are local density ridges~\citep{Platen2007,Neyrinck2008,Aragon2013}. If the density field is calculated on multiple scales or via a scale-free method such as the Voronoi tessellation~\citep{Schaap2000,vandeW2009}, the hierarchical nature of voids is naturally revealed as voids on all or many scales (limited by the simulation resolution) are found.

A study of the dynamics of voids was undertaken by~\citet{Aragon2013}. They first identify voids in the density field as the basins in a watershed transform, then calculate the velocity fields within the voids after removing the bulk flow. They found that the velocity field in void interiors is practically laminar except at the boundaries of top-level voids, and there is little shell-crossing at the boundaries of sub-voids. Again, however, the voids themselves are defined in the density field, and shell-crossing itself is not explicitly measured. 

Methods which explicitly measure shell-crossing, or find caustics, envision the gravitational collapse of dark matter as a phase-space sheet that folds on itself~\citep{Abel2012,Falck2012,Shandarin2012}. The methods of~\citet{Abel2012} and~\citet{Shandarin2012} can determine the number of caustic crossings that have occurred for each particle by counting inversions of a tessellation defined in Lagrangian space. The \org\ method of~\citet{Falck2012} only detects whether shell-crossings have occurred but keeps track of the number of orthogonal axes along which crossings have occurred to determine the cosmic web morphology of each dark matter particle. Halo particles have crossed along three axes, filaments two, walls one, and void particles are in the single-stream regime.

Other methods that characterize the dynamical cosmic web determine the three eigenvalues of either the tidal tensor (the Hessian of the gravitational potential) or the shear tensor (the Hessian of the velocity potential), giving the expansion or collapse of a volume element along three orthogonal axes~\citep{Hahn2007,Forero2009,Alonso2014,Hoffman2012,Libeskind2013,Cautun2013, Nuza2014,Metuki2014}. Note that these are measured on an Eulerian grid and thus depend on some smoothing scale, in contrast to \org.  
Studies using these methods focus on the entire cosmic web and in general produce no void catalogs, though the method of~\citet{Hahn2007} participated in the void finder comparison project of~\citet{Colberg2008}. Indeed, depending on eigenvalue threshold and smoothing scale used, methods that determine the cosmic web morphology of each volume element in a simulation are likely to produce connected or percolating super-structures~\citep{Forero2009}.

The phenomenon of percolation in large scale structure occurs when a certain type of structure -- for example, voids or filaments -- spans the volume of an $N$-body dark matter simulation such that opposite sides of the box are connected by the structure in each dimension. This has been well-studied in the density field: for a given density threshold, iso-density surfaces split up the volume into over- and under-dense regions, and the density threshold value thus determines whether the cosmic web is composed of isolated voids/clusters or a percolating super-void/super-cluster~\citep{Shandarin2004,Shandarin2006,Shandarin2010}. In cosmic web methods, instead of a density threshold, percolation is a function of the eigenvalue threshold, $\lambda_{th}$, for collapse; in principle this would be zero, but such a threshold does not produce a cosmic web that matches the visual impression in the density field~\citep{Forero2009}. There is some transition threshold above or below which percolation occurs, but these also depend on scale.
\org, on the other hand, explicitly determines whether shell-crossing has occurred for every particle in an $N$-body simulation and introduces no added smoothing scale. It is thus well-suited to the task of determining whether void boundaries are defined by shell-crossing, i.e., addressing whether shell-crossing prevents the percolation of single-stream regions.

This paper investigates the nature of single-stream voids, which are defined as regions that have not undergone any shell-crossing. In particular, we study the percolation properties of \org\ void particles. Since \org\ applies no smoothing scale, the relevant scale is the simulation resolution; as resolution increases, more shell-crossing is detected, resulting in a higher mass fraction of halo particles~\citep{Falck2012} as a result of the power spectrum of density fluctuations in cold dark matter (CDM). In Section~\ref{sec:sims} we describe the suite of simulations used for this study. We describe the \org\ algorithm in Section~\ref{sec:origami} and show mass and volume fractions of the cosmic web at multiple resolutions. Our single-stream percolation results are presented in Section~\ref{sec:singleperc}, and we measure the percolation of multi-stream particles in Section~\ref{sec:multiperc}. We address potential limits to shell-crossing detection in Section~\ref{sec:shelleff} and speculate on the continuum limit of infinite resolution in Section~\ref{sec:highres}. Finally, in Section~\ref{sec:voiddef} we define a catalog of \org\ voids by introducing a volume threshold parameter, and we investigate how the simulation resolution and volume parameter affect the percolation of these single-stream voids.


\section{Methods}
\label{sec:methods}

In this section, we first describe the $N$-body simulations used in Section~\ref{sec:sims}. We then briefly describe the \org\ algorithm for identifying whether, and in how many dimensions, dark matter particles have undergone shell-crossing in Section~\ref{sec:origami}; we refer the reader to~\citet{Falck2012} for more details on the algorithm. Here we focus on the resolution dependence of the mass and volume fractions of the different cosmic web elements. Finally, in Section~\ref{sec:percstats} we describe our method of measuring percolation properties in these simulations.

\subsection{Simulations}
\label{sec:sims}

We use a suite of six Gadget-2~\citep{Springel2005} $N$-body dark matter simulations. There are two box sizes, 100\hmpc\ and 200\hmpc; for each box size, the same initial conditions are simulated with three different particle numbers ($128^3$, $256^3$, and $512^3$), where the lower resolution versions are down-sampled from the higher-resolution.\footnote{We note that while these may not be very large volume simulations, the agreement between them (at the same resolution) in the following sections, though they have different initial phases, is encouraging that our results are not affected by cosmic variance.} They are simulated with a standard $\Lambda$ cold dark matter (LCDM) cosmological model using Planck parameters: $h=0.68$, $\Omega_M = 0.31$, $\Omega_\Lambda = 0.69$, $n_s = 0.96$, and $\sigma_8 = 0.82$. This gives dark matter particle masses of $3.25\times 10^{11}$, $4.07\times 10^{10}$, $5.08\times 10^9$, and $6.35\times 10^8$ \hmsun\ from the lowest to highest resolution. 

Throughout this paper, we will refer to specific simulations by their inter-particle separation, $L/N$; these range from 0.2\hmpc\ at the highest resolution to 1.6\hmpc\ at the lowest resolution, with two simulations each at $L/N=0.4$\hmpc\ and $L/N=0.8$\hmpc. This initial inter-particle separation provides a physical scale with which to interpret the results, since there is no other smoothing.  Note that at late times, particles in high-density regions will be much closer to each other than $L/N$, so the density field (and \org\ cosmic web) will be sampled on much smaller scales, but the density field in voids will be sampled on scales $\gtrsim L/N$ since particles in voids move very little.

We additionally use two simulations with slightly different cosmological parameters to see whether the quantities we calculate depend on cosmology. The LCDM parameters for these are WMAP-era, and they will be referred to in what follows as the WMAP simulations to distinguish them from the Planck simulations described above. These parameters are: $h=0.73$, $\Omega_M = 0.3$, $\Omega_\Lambda = 0.7$, $n_s = 0.93$, and $\sigma_8 = 0.81$. Note that these simulations are completely separate from the Planck simulations and have different initial Fourier phases. Both WMAP simulations have a box length of 100\hmpc\ and the same initial conditions, but the lower resolution ($256^3$ particle, $L/N = 0.4$\hmpc) simulation has been down-sampled from the higher resolution ($512^3$ particle, $L/N = 0.2$\hmpc) simulation in the initial conditions. The dark matter particle masses for these are $4.45\times 10^9$ and $5.58\times 10^8$ \hmsun\ for the lower and higher resolution, respectively.


\subsection{ORIGAMI}
\label{sec:origami}

As particles in a dark matter simulation collapse under gravity to form structures, their trajectories cross and enter a nonlinear phase of shell-crossing. Another way to think of this process is to imagine that the particles are the vertices of an initially regular grid, or a three-dimensional dark matter sheet. Gravitational collapse causes the initially flat sheet to distort and stretch; in six-dimensional phase-space, this three-dimensional sheet folds without intersecting itself when shell-crossing occurs. These folds mark out the boundaries of the collapsing structures -- additionally, the dimensionality of the folding determines the structure morphology: one-dimensional collapse forms a wall, two-dimensional collapse forms a filament, and three-dimensional collapse forms a halo. \org\footnote{\url{http://icg.port.ac.uk/~falckb/origami.html}} identifies the shell-crossing by taking advantage of the fact that the relative positions of particles within folded regions have reversed with respect to their initial locations on the Lagrangian grid.

\org\ tags particles that have crossed along 0, 1, 2, and 3 orthogonal axes as void, wall, filament, and halo particles, respectively. This number is a particle's \org\ morphology index $M$. To make sure relevant crossings aren't missed, \org\ looks for particle reversals along four sets of orthogonal axes (see~\citet{Falck2012} for details). The morphology index $M$ returned for each particle is the maximum $M$ found among all four sets of axes. Note that $M$ is calculated by comparing the particle positions at a given snapshot to their  Lagrangian positions on the grid; no information from previous snapshots is used, nor is it needed to catch the overwhelming majority of crossings (see Section~\ref{sec:shelleff} for details).

\begin{figure*}
\includegraphics[width=\hsize]{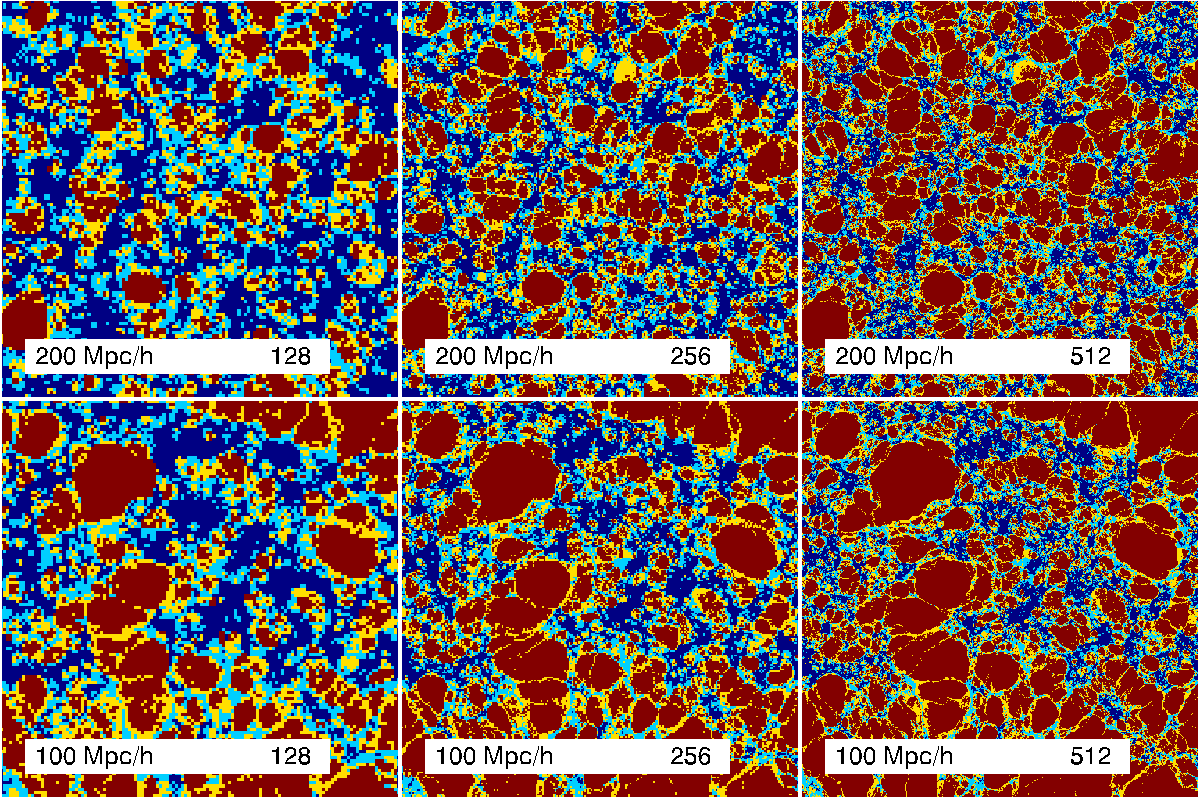}
\caption{Slices of redshift 0 \org\ morphology of particles placed at their Lagrangian positions on the initial grid, for each of the six Planck simulations, identified by their box size and 1-D particle number. Halo particles are in red, filament in yellow, wall in light blue, and void in dark blue; the upper left panel is the lowest resolution, the bottom right is the highest, and the 100 and 200\hmpc\ boxes have been scaled to have the same size. Halos in Lagrangian space show up as blobs, surrounded by filament and wall particles. As resolution increases, void regions fill up with small-scale structures.
\label{fig:mlag}}
\end{figure*}

Because \org\ has no added scale or smoothing, the relative numbers of particles with a given $M$ in a simulation, i.e. the fractions of halo, filament, wall, and void particles, depend on the simulation resolution. As resolution increases and smaller structures are detected, the average $M$ in a given Lagrangian region tends to increase, resulting in a higher fraction of halo particles and a lower fraction of void particles overall. Figure~\ref{fig:mlag} shows the \org\ morphology of particles from the suite of six Planck simulations, where the 200\hmpc\ box has been scaled down to the same size as the 100\hmpc\ box. The redshift 0 morphologies of particles are shown at their Lagrangian positions such that each pixel in the image is one particle. The upper left panel is the lowest resolution simulation with $L/N=1.6$\hmpc, and the lower right panel is the highest resolution simulation with $L/N=0.2$\hmpc. In this Lagrangian view, halos at $z=0$ show up as blobs, bordered by filament and wall particles. As resolution increases, regions of void and wall particles become filled with smaller regions of filament and halo particles.

\begin{figure}
\includegraphics[angle=-90,width=\hsize]{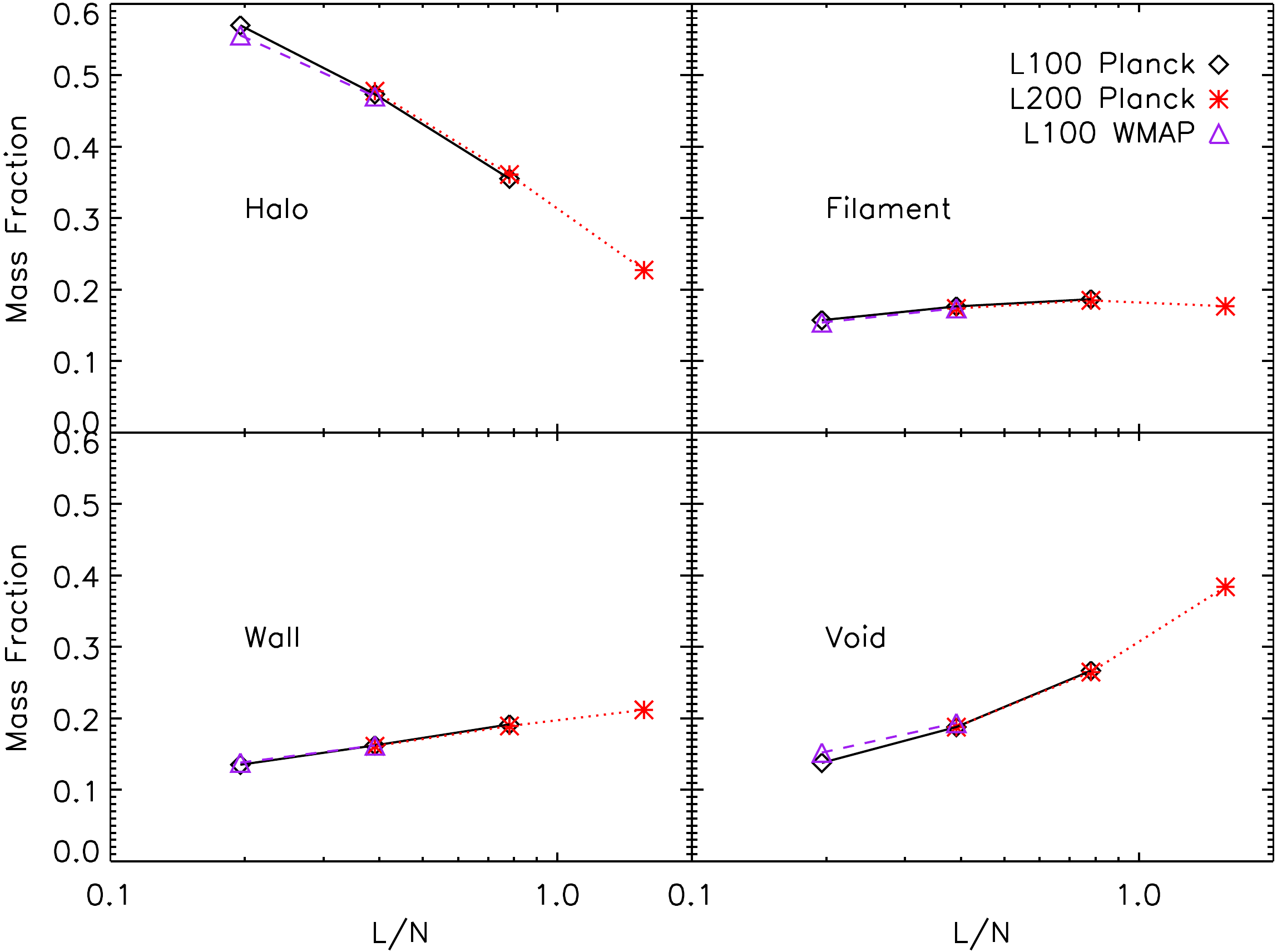}
\caption{Particle mass fractions as a function of simulation resolution and \org\ morphology. Increasing the simulation resolution (lowering $L/N$) increases the fraction of halo mass and decreases the fraction of void mass, and there is very good agreement between simulations with the same resolution but different initial conditions.
\label{fig:mfracs}}
\end{figure}

We can quantify this picture with the mass fractions (or equivalently, the fractions of the number of particles) of the different \org\ cosmic web morphologies. These are plotted as a function of resolution, given by the initial particle separation $L/N$, in Figure~\ref{fig:mfracs}, for all six Planck simulations and the two WMAP simulations. There is very good agreement between the two sets of Planck simulations that have the same resolution, whereas there is some slight offset between the halo and void mass fractions of the WMAP and Planck simulations sharing the same resolution. The fraction of halo particles increases quite dramatically as simulation resolution increases, and the fraction of void particles decreases equally dramatically, while the filament and wall mass fractions do not vary as much. 

Another important quantity to consider is the volume fraction of each cosmic web element. We calculate the volume and density of each particle using the Voronoi Tessellation Field Estimator~\citep[VTFE,][]{Schaap2000,vandeW2009,Pandey2013}.  The Voronoi density for each particle is given by $\delta = \bar{V}/V - 1$, where $V$ is the volume of the particle's surrounding Voronoi cell and $\bar{V}$ is the average of all cell volumes. The Voronoi tessellation and its dual, the Delaunay tessellation, also provide a set of nearest neighbors for each particle. The neighbors of a particle are given by the set of adjacent Voronoi cells, or equivalently, the vertices of the same Delaunay tetrahedron. We note here that a density estimate based on the tessellation of particles in Lagrangian space (LTFE) would provide different and possibly less biased densities, especially around $\delta=1$, as noted by~\citet{Abel2012}; however, the LTFE density is likely biased high in the centers of halos and low just outside halo centers because of phase-space overwindings~\citep{Hahn2013,Angulo2014}. In any case, our results on the percolation of single- and multi-stream regions in the next section are independent of density estimator.

\begin{figure}
\includegraphics[angle=-90,width=\hsize]{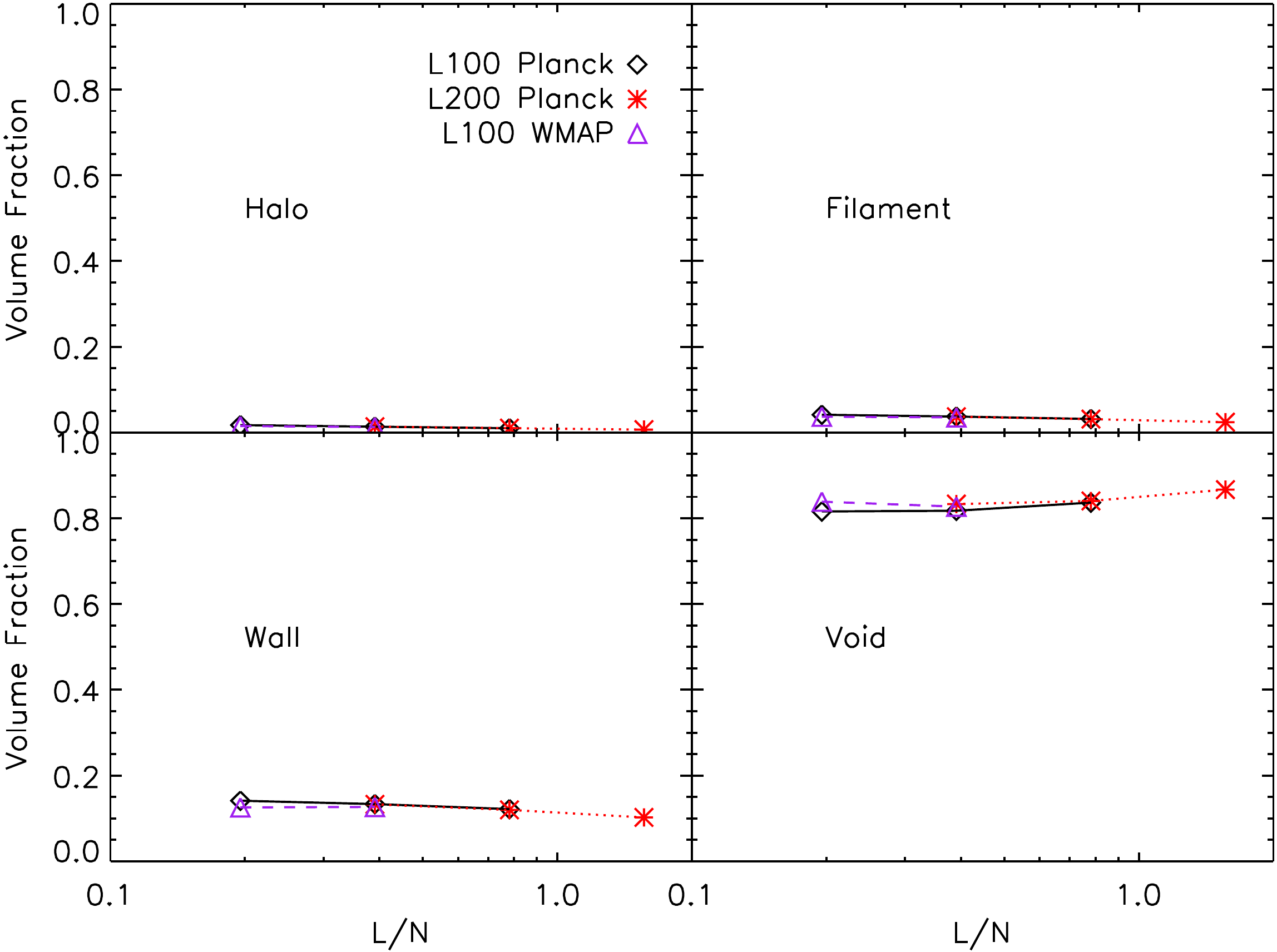}
\caption{Particle volume fractions as a function of simulation resolution and \org\ morphology. Single-stream voids dominate the volume at all resolutions, and the void volume fraction only decreases slightly as resolution increases, remaining above 80\%.
\label{fig:vfracs}}
\end{figure}

In Figure~\ref{fig:vfracs} we show the volume fractions as a function of simulation resolution for each type of cosmic web morphology. The most striking feature, comparing this to Figure~\ref{fig:mfracs}, is that the dependence on $L/N$ is very weak for the volume fractions while strong for the mass fractions. As resolution increases and a higher fraction of collapsed mass is resolved, these newly resolved structures add only very slightly to the total volume in collapsed structures. Again there is very good agreement between the Planck simulations with the same resolution, and small offsets between those with the same resolution but different cosmological parameters. 

\begin{figure}
\includegraphics[angle=-90,width=\hsize]{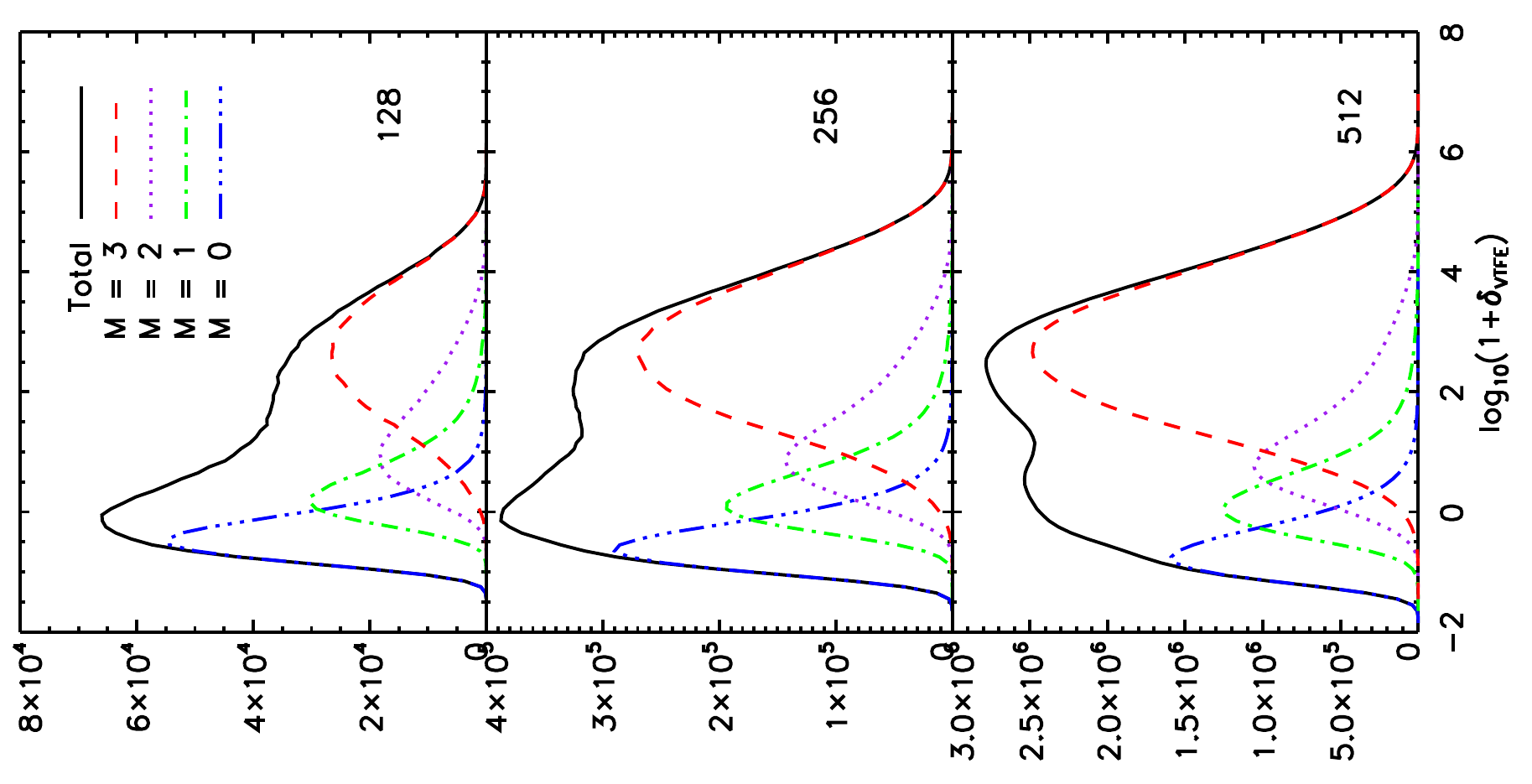}
\caption{VTFE log-density distribution functions of particles in the three 100\hmpc\ Planck simulations as a function of \org\ morphology. Void particles ($M=0$) are in blue, wall ($M=1$) in green, filament ($M=2$) in purple, and halo ($M=3$) in red. 
\label{fig:rhotag}}
\end{figure}

Though halos have the highest densities and voids the lowest, there is significant overlap in the density ranges of the different morphological structures~\citep{Hahn2007,Aragon2010,Falck2012}. This implies that a density threshold is not sufficient to distinguish between the different morphological types. Figure~\ref{fig:rhotag} shows the probability distribution functions of the particle densities split according to their \org\ morphological classification for all three 100\hmpc\ Planck simulations. The halo particles create a high-density bump in the density distribution which grows as simulation resolution increases and there are correspondingly more high-density particles. Individually, the density distributions of each morphological type appears to be close to lognormal. We know of no particular theoretical motivation for the functional form of these PDFs, though an approximately lognormal analytic distribution has been found for void particles that works reasonably well~\citep{Neyrinck2013}.

\begin{figure}
\includegraphics[angle=-90,width=\hsize]{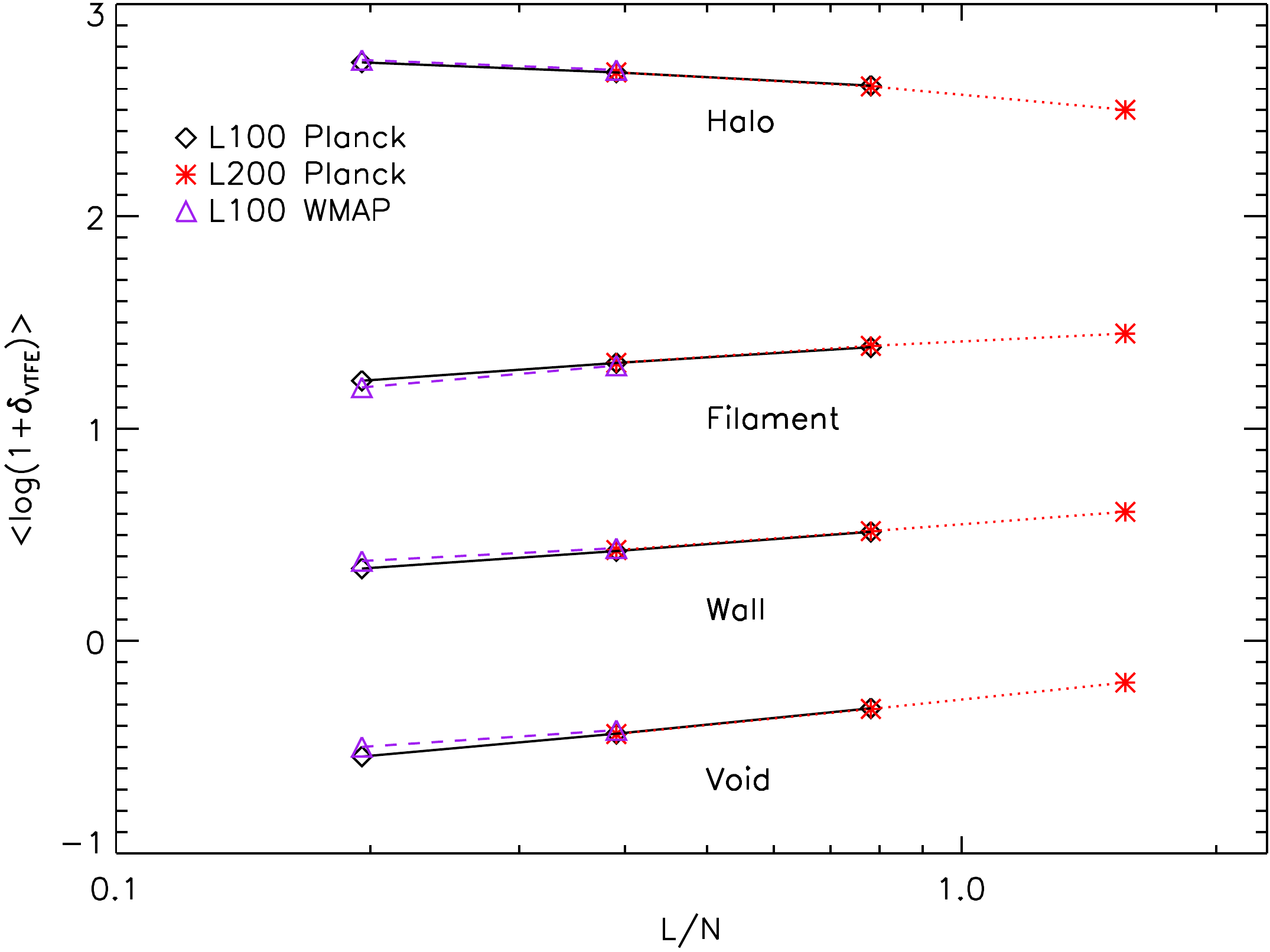}
\caption{Average VTFE density of halo, filament, wall, and void particles as a function of simulation resolution. As resolution increases, the average density of halo particles increases while the average density of wall, filament, and void particles decreases.
\label{fig:avgdens}}
\end{figure}

The average VTFE densities, $\langle\log(1+\delta_{\rm VTFE})\rangle$, of halo, filament, wall, and void particles are plotted as a function of simulation resolution in Figure~\ref{fig:avgdens}, for all simulations. As the resolution increases and a greater fraction of halo particles and lower fraction of void particles are identified, the average VTFE density of filament, wall, and void particles decreases slightly while the average density of halo particles increases. The relative fractions of particles with different morphologies, e.g. the fraction of halo particles and height of the high-density bump in Figure~\ref{fig:rhotag}, also depend on the cosmological model~\citep{Falck2014}, in addition to the slight dependence on WMAP vs. Planck LCDM parameters.

We stress again that both the \org\ cosmic web identification and VTFE density field are independent of any added parameter or smoothing scale. The important quantity is thus the simulation resolution, referred to here by the inter-particle separation $L/N$. In Section~\ref{sec:highres}, we speculate on how the mass and volume fractions of the collapsed matter and the single-stream regions might extend to infinite resolution based on these plots, with a view to what that might mean for the percolation of single-stream voids.


\subsection{Quantifying Percolation}
\label{sec:percstats}

Previous studies have looked at percolation as a function of both smoothing scale and the value of either a density threshold~\citep{Shandarin2004, Shandarin2006, Shandarin2010} or eigenvalue collapse threshold~\citep{Forero2009}. In these studies, connectivity is defined on a real space grid. Since \org\ introduces no smoothing scale, however, the relevant scale is given by the simulation resolution; this scale also affects the classification of dark matter particles into the different components of the cosmic web, but there is no threshold parameter. 

To measure the percolation of a set of particles, we iteratively connect particles that are neighbors in the Voronoi/Delaunay tessellation. This is like a friends-of-friends algorithm but with no linking length -- particles connected on the tessellation are very close together in high-density regions and far apart in low-density regions. Again, as with the VTFE density and \org\ morphology, this gives a definition of connectivity that is independent of a (smoothing) scale, depending only on simulation resolution. This is quite different than grid-based studies of percolation which connected only the six nearest neighbors of cells in a cubic mesh~\citep{Shandarin2004,Forero2009,Shandarin2010}; by contrast, the average number of tessellation neighbors for Poisson-distributed particles is $\approx 15.5$~\citep{Lazar2014}. However, note that for a cubic mesh, the six nearest grid cells completely encompass the neighboring {\it volume} of a given cubic cell, as does the set of neighbors in a Voronoi/Delaunay tessellation.

Once we have connected single-stream (or multi-stream) particles on the tessellation, this gives us a catalog of groups of connected particles. To determine whether percolation has occurred, we calculate the mass and volume fractions of the largest structure with respect to all particles of that type. Taking \org\ void particles as an example, the mass fraction would be the number of void particles in the largest void divided by the total number of void particles, or $N_{\rm max}/N_{\rm void}$, and the percolating volume fraction would be the total volume (given by the sum of the VTFE volumes) in the largest void divided by the volume in all voids, or $V_{\rm max}/V_{\rm void}$. For example, ~\citet{Forero2009} find that $V_{\rm max}/V_{\rm void}$ jumps from 0.1 to 0.9 very quickly as a function of the eigenvalue threshold used to define the different components of the cosmic web, for a fixed smoothing scale. \org\ has no parameter to define the cosmic web and no added smoothing scale, but since the cosmic web identification depends on the simulation resolution itself (see Figures~\ref{fig:mfracs} and~\ref{fig:vfracs}), we will determine how the percolating mass and volume fractions change with the inter-particle separation $L/N$.

The mass and volume fractions of the largest structure are not always so large that the structure is obviously percolating; this is especially true when we include a density parameter in Section~\ref{sec:voiddef}. To determine whether percolation has occurred in these borderline cases, we measure the minimum, maximum, and average particle position of the largest structure along the $x$, $y$, and $z$ directions. A percolating structure will extend from 0 to $L$ along all three axes of the box and have average particle positions of roughly $L/2$.


\section{Persistent Percolation}
\label{sec:percolation}

In this section, we show that shell-crossing does not split voids into distinct, individual regions; rather, the single-stream voids percolate, and this percolation is persistent. By this we mean that no transition to non-percolating single-stream regions is found despite our efforts, which include increasing the simulation resolution to increase the fraction of multi-stream regions, limiting the considered single-stream particles to only those that are not connected to any multi-stream particles, and searching for deficiencies in the shell-crossing detection of our algorithm.
We measure and quantify the percolation of both single- and multi-stream regions, and subsets of these regions, as a function of simulation resolution in Sections~\ref{sec:singleperc} and~\ref{sec:multiperc}. In Section~\ref{sec:shelleff} we show that while the detection of shell-crossing itself is not perfect and a small fraction of crossings can be missed by the algorithm, this is likely not enough to affect percolation properties. Finally, in Section~\ref{sec:highres} we discuss the limit of infinite resolution, and we speculate on whether the fraction of collapsed mass will ever be high enough to prevent single-stream percolation.

\subsection{Single-stream Percolation}
\label{sec:singleperc}

\begin{figure}
\includegraphics[angle=-90,width=\hsize]{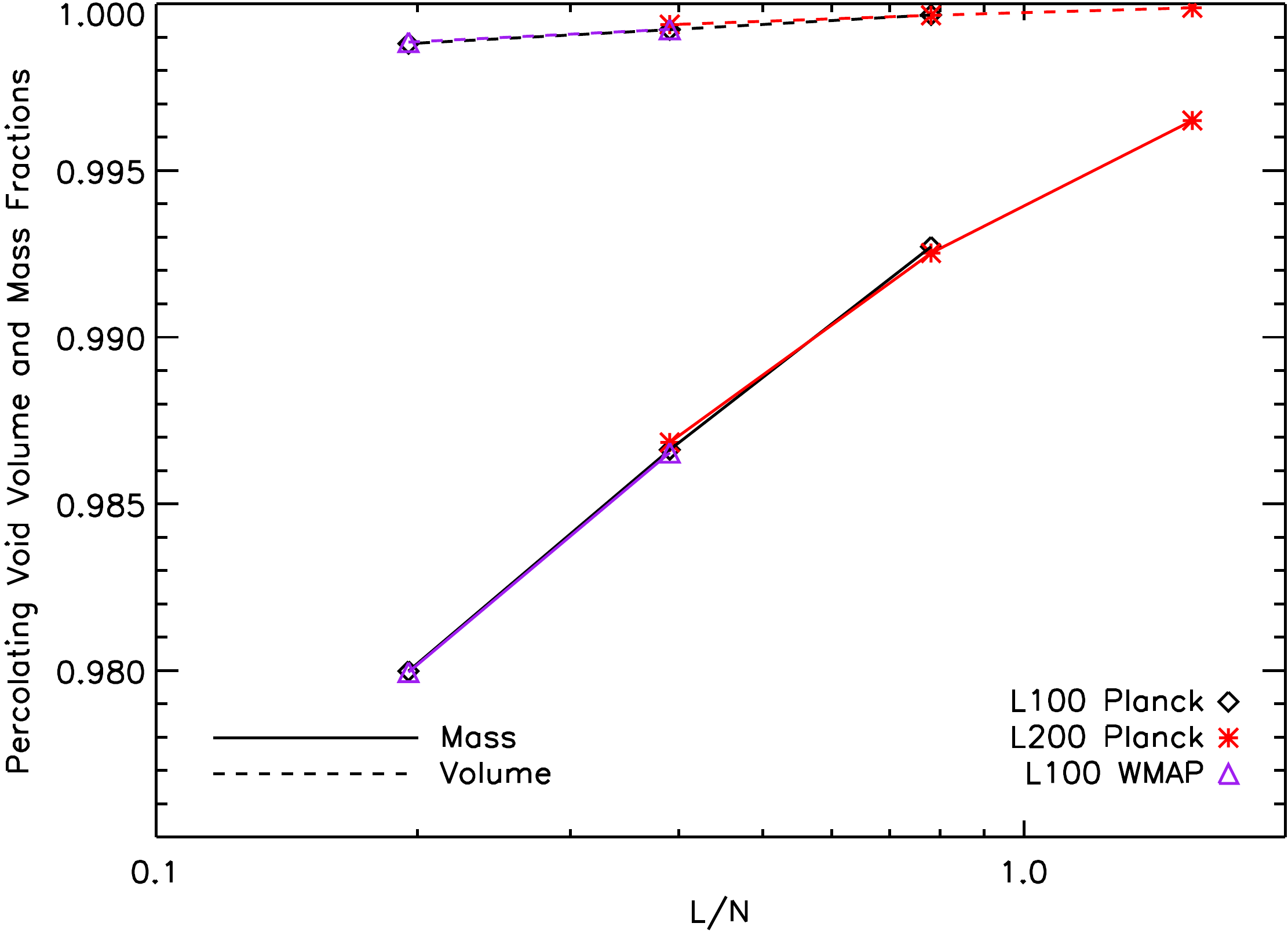}
\caption{Mass and volume fractions of the largest (percolating) void, where fractions are with respect to the mass and volume of all void particles. Percolation persists down to the highest resolution studied here, with 98\% of the void particles in the percolating void.
\label{fig:voidperc}}
\end{figure}

We start with single-stream voids. As described in Section~\ref{sec:methods}, we group together \org\ void particles that are neighbors on the Delaunay tessellation to determine the connectivity of the single-stream regions. It turns out that the largest single-stream structure percolates the volume at all simulation resolutions. The mass and volume fractions of the percolating void are given in Figure~\ref{fig:voidperc} for all eight simulations as a function of resolution, $L/N$. The overwhelming majority of the void mass and volume is in the largest void at all resolutions; clearly, the filament, wall, and halo particles do not divide the set of single-stream void particles into distinct, individual voids. The volume fractions are all above 99.8\%, with only a slight decrease as resolution increases; the mass fractions go from 99.7\% at the lowest resolution to 98\% at the highest resolution.
As with the mass and volume fractions of the cosmic web elements given in Section~\ref{sec:origami}, there is again remarkable agreement between simulations with different initial conditions (and box sizes) and the same resolution, and between those with WMAP and Planck cosmological parameters.

Thus, though increasing the resolution of the simulation increases the fraction of collapsed structures detected, it does little to reduce the size of the percolating void. These new structures detected at higher resolutions are small-scale halos, filaments, and walls, adding significantly to the mass fractions but little to the volume fractions; from Figures~\ref{fig:mfracs} and~\ref{fig:vfracs}, the mass fraction of all void particles reduces from 40\% at the lowest resolution to 14\% at the highest, while the volume fraction only reduces from 85\% to 80\%. Note that in previous studies of percolation in the density field~\citep{Shandarin2006} and in the tidal tensor description of the cosmic web~\citep{Forero2009}, percolation begins when the void volume fraction is much lower (20-25\%). It is doubtful that single-stream voids will ever reach such low volume fractions even as resolution is increased arbitrarily, judging by the trend in Figure~\ref{fig:vfracs}. This high void volume fraction suggests that single-stream percolation may persist to infinite resolution; we will return to this subject in Section~\ref{sec:highres}.
The percolating mass fractions in Figure~\ref{fig:voidperc} also show a greater variation with resolution than the volume fractions. As resolution increases, the new resolved structures eat away at the edges and insides of the percolating void and leave more isolated, high-density voids (\textit{void-in-cloud}), while most of the single-stream void volume remains connected.

To reduce the chance that percolation is occurring through ``holes'' in walls that divide single-stream voids, we measure the percolation of a subset of void particles that are neighbors only of other void particles. Specifically, we only connect void particles that have a fraction of void neighbors, $f_v$, equal to unity, thus ignoring void boundaries that are connected to collapsed structures on the tessellation. For the Planck simulations, 22\% to 25\% of void particles have $f_v=1$ (depending on simulation resolution), corresponding to 36\% to 42\% of the void volume, so we are considering a significantly reduced portion of single-stream voids which constitute their inner regions. On average, 71\% to 77\% of the tessellation neighbors of void particles are other void particles. The fraction of void neighbors, $f_v$, increases by a small amount as resolution decreases (since the void volume and mass fractions are higher) for all morphology types. 

\begin{figure}
\includegraphics[angle=-90,width=\hsize]{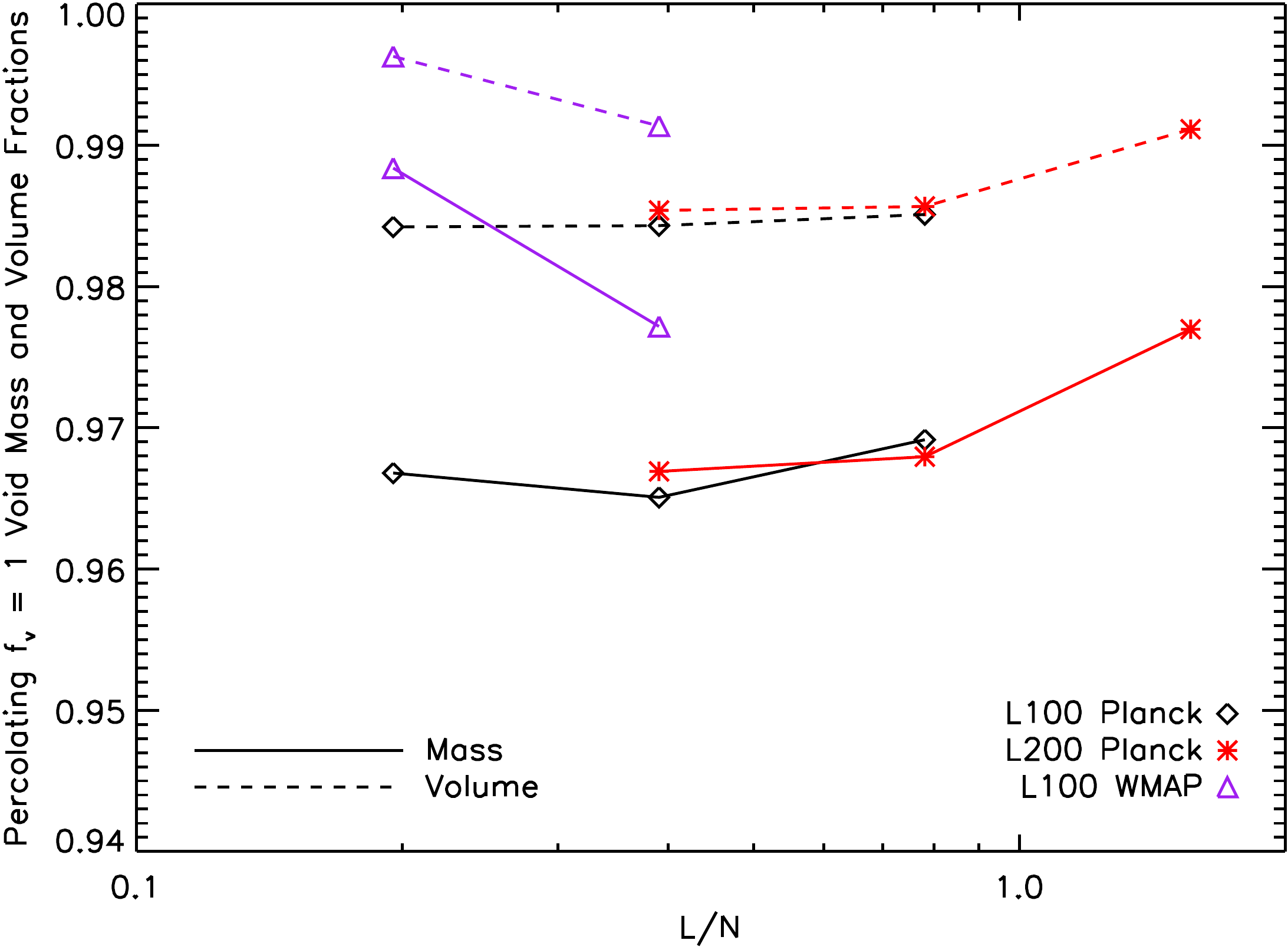}
\caption{Mass and volume fractions of the largest (percolating) $f_v=1$ void, consisting only of void particles that are completely surrounded by other void particles in the tessellation, thereby avoiding boundaries of single-stream regions. This subset of void particles clearly percolates at all simulation resolutions, thus single-stream percolation is not likely caused by ``holes'' in walls. 
\label{fig:vf1perc}}
\end{figure}

All of the simulations have $f_v=1$ percolating void mass and volume fractions above 96\%, shown in Figure~\ref{fig:vf1perc}. The percolating fractions do not have the same resolution dependence as the previous case of all single-stream particles and do not show the same agreement between different simulations at the same $L/N$, but since the restriction of $f_v=1$ does not have an obvious physical interpretation, the details are likely unimportant.
The important thing to note is that the set of $f_v=1$ void particles by construction contains the inner regions of all voids. Though this percolating void is much reduced in both particle number and size as compared to the set of all void particles, it seems that excluding the boundaries of single-stream regions does not prevent percolation. 

It may seem odd that the mass and volume fractions of the percolating structure either levels off or increases as resolution increases in Figure~\ref{fig:vf1perc}, but for $f_v=1$ voids, increasing the simulation resolution likely increases the size of the (boundary-excluded) percolating void because the boundary itself, consisting of $f_v < 1$ void particles, gets thinner. Thus, even being conservative about the edges of the single-stream regions by only considering the connectivity of $f_v = 1$ void particles, the single-stream regions percolate the volume. 

\begin{figure}
\includegraphics[angle=-90,width=\hsize]{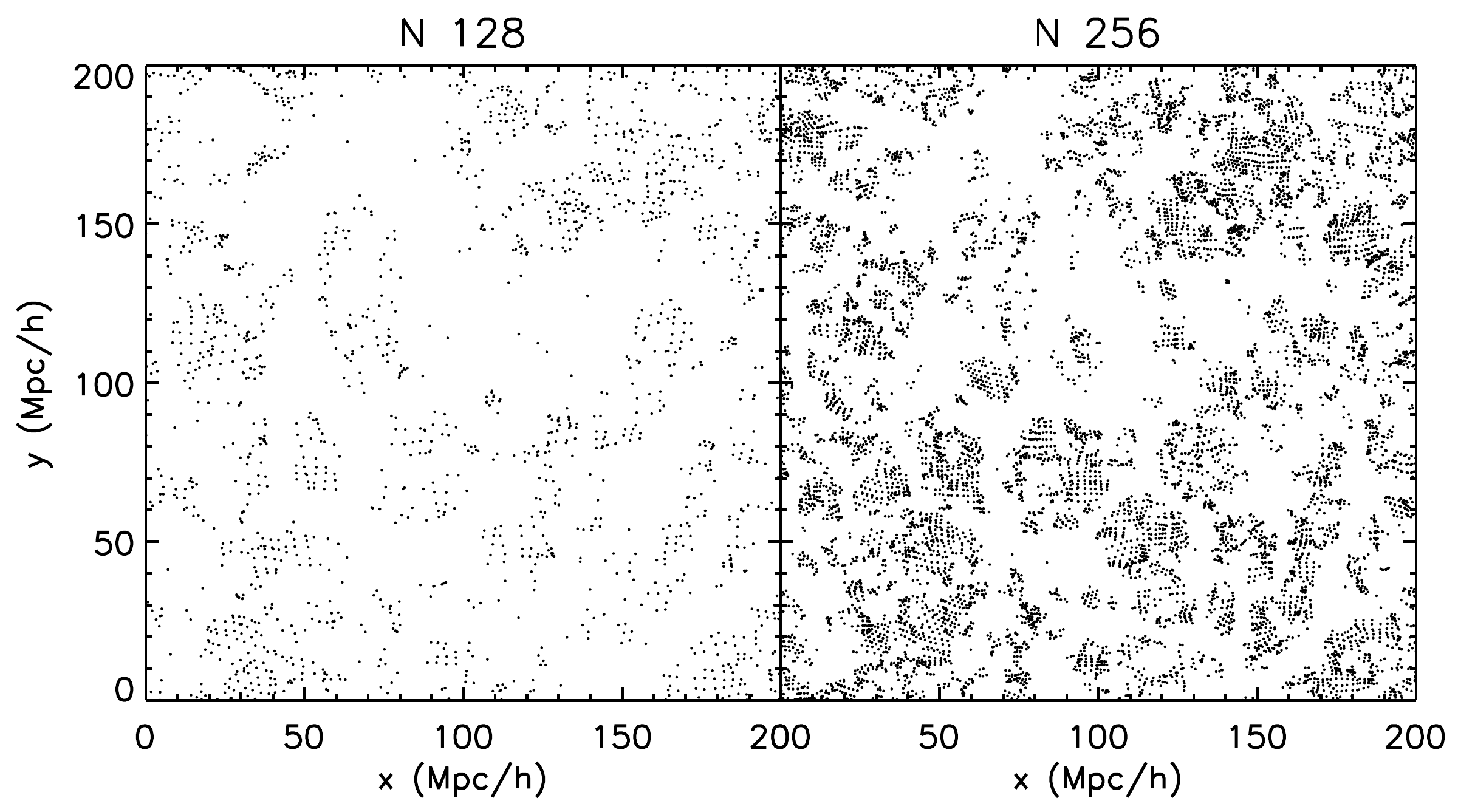}
\caption{A thin, 2\hmpc-thick slice through the 200\hmpc\ Planck simulations with $L/N=1.6$\hmpc\ (left) and 0.8\hmpc\ (right), showing only particles in the percolating $f_v=1$ void. Even this subset of single-stream regions that avoid multi-stream boundaries, consisting of 22\% to 25\% of void particles, spans the volume and percolates.
\label{fig:slice_vf1}}
\end{figure}

Since the percolating single-stream void contains the overwhelming majority of void particles, it is perhaps not very useful to visualize. Instead, we show the network of the largest percolating structure of $f_v = 1$ void particles in Figure~\ref{fig:slice_vf1}, for two resolutions of the 200\hmpc\ Planck simulations. Both slices are 2\hmpc\ thick, so the higher-resolution slice on the right contains more particles than the lower-resolution slice on the left, but in both, these innermost single-stream particles span the slice and obviously percolate the three-dimensional volume.

Single-stream percolation is thus persistent. The largest single-stream structure percolates even down to a resolution of $L/N$ = 0.2\hmpc, when only about 14\% of the mass is in the single-stream regime, and this is independent of small changes to LCDM cosmological parameters. This single-stream region contains about 80\% of the volume which changes very little as resolution increases, so even at very high resolutions when a small fraction of the mass is in the single-stream regime, single-stream voids will likely continue to percolate. Some discussion of the limit of infinite resolution is given in Section~\ref{sec:highres}. 

Single-stream percolation also persists when considering only the innermost void regions, defined by void particles that are only neighbors of other void particles on the tessellation so that their fraction of void neighbors, $f_v$, is unity. Though these make up only 22\% to 25\% of void particles and 36\% to 42\% of the void volume (depending on resolution), they nevertheless create a percolating structure; there simply aren't enough wall and filament regions to divide the single-stream regions into individual, isolated voids. Since individual, isolated voids obviously do exist when defined by the density field~\citep[e.g.,][]{Neyrinck2008,Aragon2013}, the density ridges that separate voids have not undergone shell-crossing, at least not on all sides of the void.

\subsection{Multi-stream percolation}
\label{sec:multiperc}

\begin{figure}
\includegraphics[angle=-90,width=\hsize]{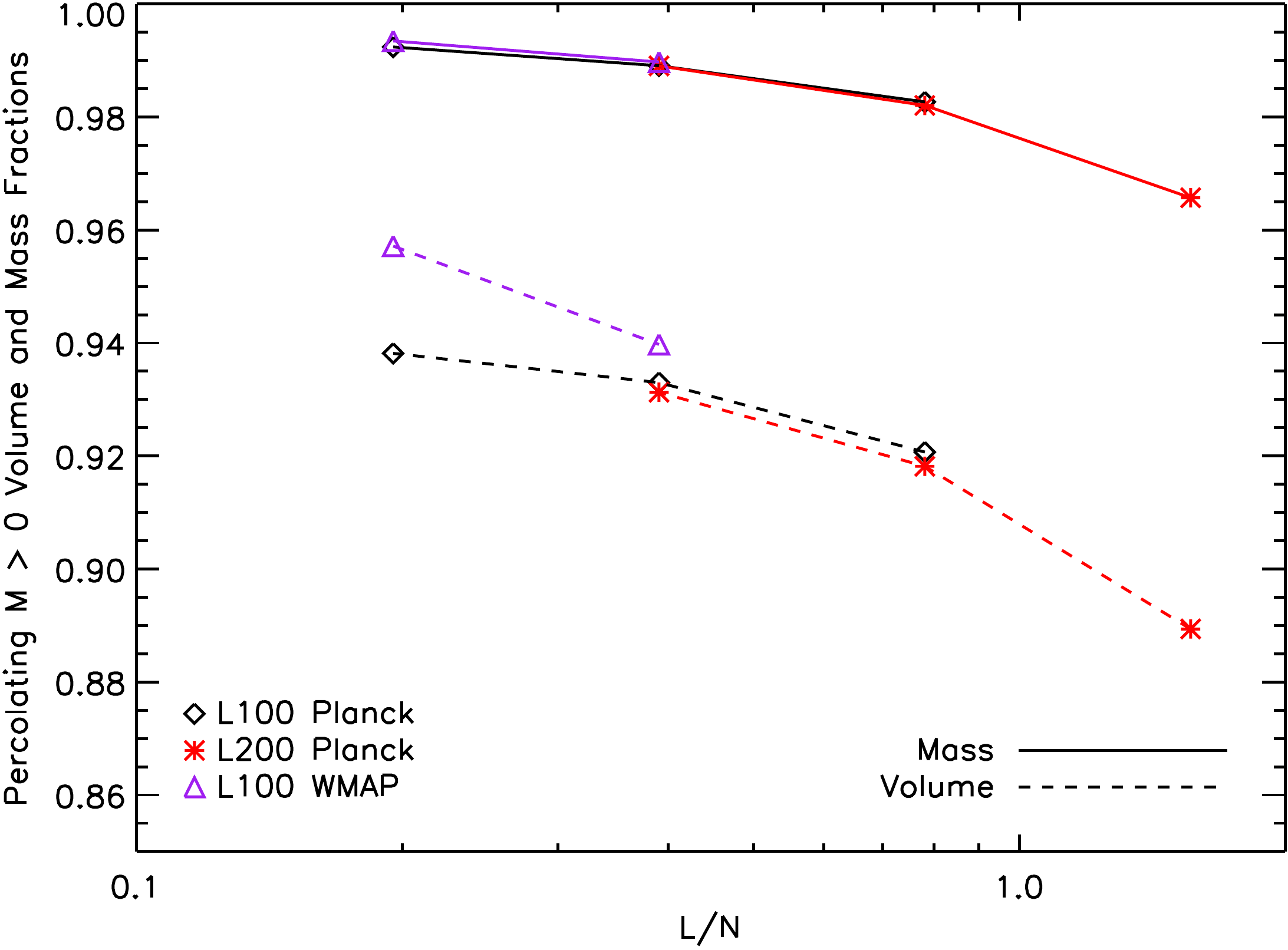}
\caption{Mass and volume fractions of the percolating multi-stream structures (i.e., particles with \org\ morphology index $M > 0$) as a function of resolution, where fractions are with respect to the mass and volume of all $M>0$ particles. The trend with resolution is opposite that of single-stream percolation.
\label{fig:m1perc}}
\end{figure}

The set of all multi-stream regions containing wall, filament, and halo particles ($M>0$) also creates a percolating structure. Again, $M$ is the \org\ morphology index and gives the number of axes along which shell-crossing has occurred for an individual particle. The mass and volume fractions of the largest $M>0$ structure, with respect to all $M>0$ particles, are given in Figure~\ref{fig:m1perc} as a function of $L/N$. The trend with resolution is opposite to that of the single-stream percolation in Figure~\ref{fig:voidperc}, and the mass and volume fractions are lower, but even at the lowest resolution the volume fraction of the percolating multi-stream structure is above 88\%; even though most of the volume is in the single-stream regime at this resolution (above 85\%; see Figure~\ref{fig:vfracs}), the set of all multi-stream particles connects into a percolating structure. Given that a large range of density thresholds leads to a percolating supercluster-void network~\citep{Shandarin2004}, it is perhaps not surprising that both the single-stream and multi-stream regions form percolating structures when the boundary is defined by shell-crossing instead of a density threshold. 

\begin{figure}
\includegraphics[angle=-90,width=\hsize]{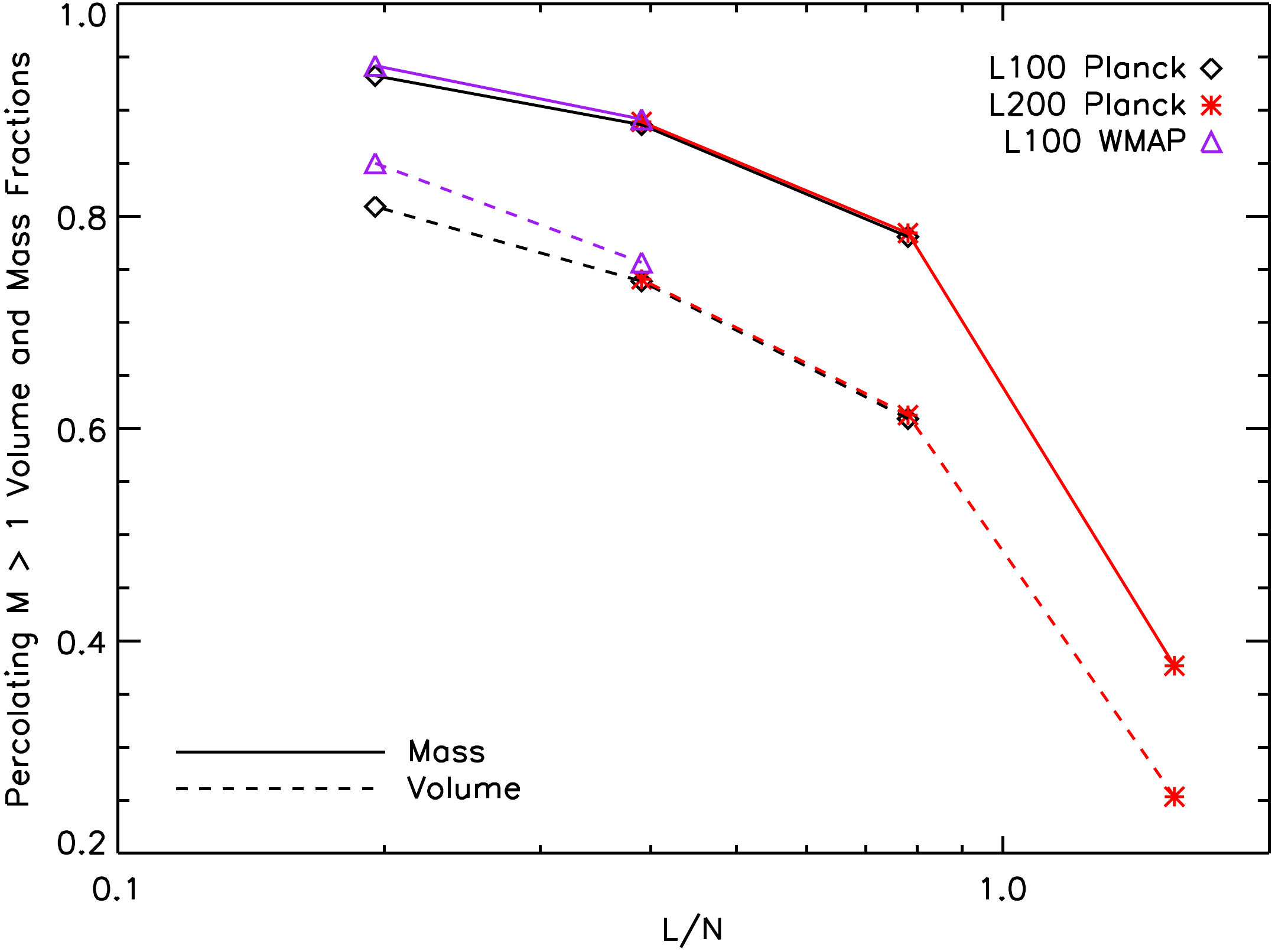}
\caption{Mass and volume fractions of the largest network of $M>1$ (halo and filament) particles, as a function of resolution. There is greater variation than found for single-stream and multi-stream percolation, but even at the lowest resolution studied here, the largest structure spans the simulation volume in all three dimensions.
\label{fig:m2perc}}
\end{figure}

Continuing to higher ``dimensions'' of shell-crossing, the network of halos and filaments ($M>1$, excluding walls and voids) also connects into a percolating structure at all simulation resolutions, but there is a drastic variation in the largest structure's mass and volume fractions as resolution decreases, shown in Figure~\ref{fig:m2perc}. At the highest resolution, the percolating $M>1$ structure contains over 90\% of all halo and filament particles and 80\% of the halo and filament volume. At the lowest resolution, these decrease to 40\% of the mass and 25\% of the volume, putting into doubt whether the structure actually percolates. We thus measure the extent and average particle positions of the largest structure along all three axes (as described in Section~\ref{sec:percstats}) and confirm that it does indeed span the simulation volume.

\begin{figure}
\includegraphics[angle=-90,width=\hsize]{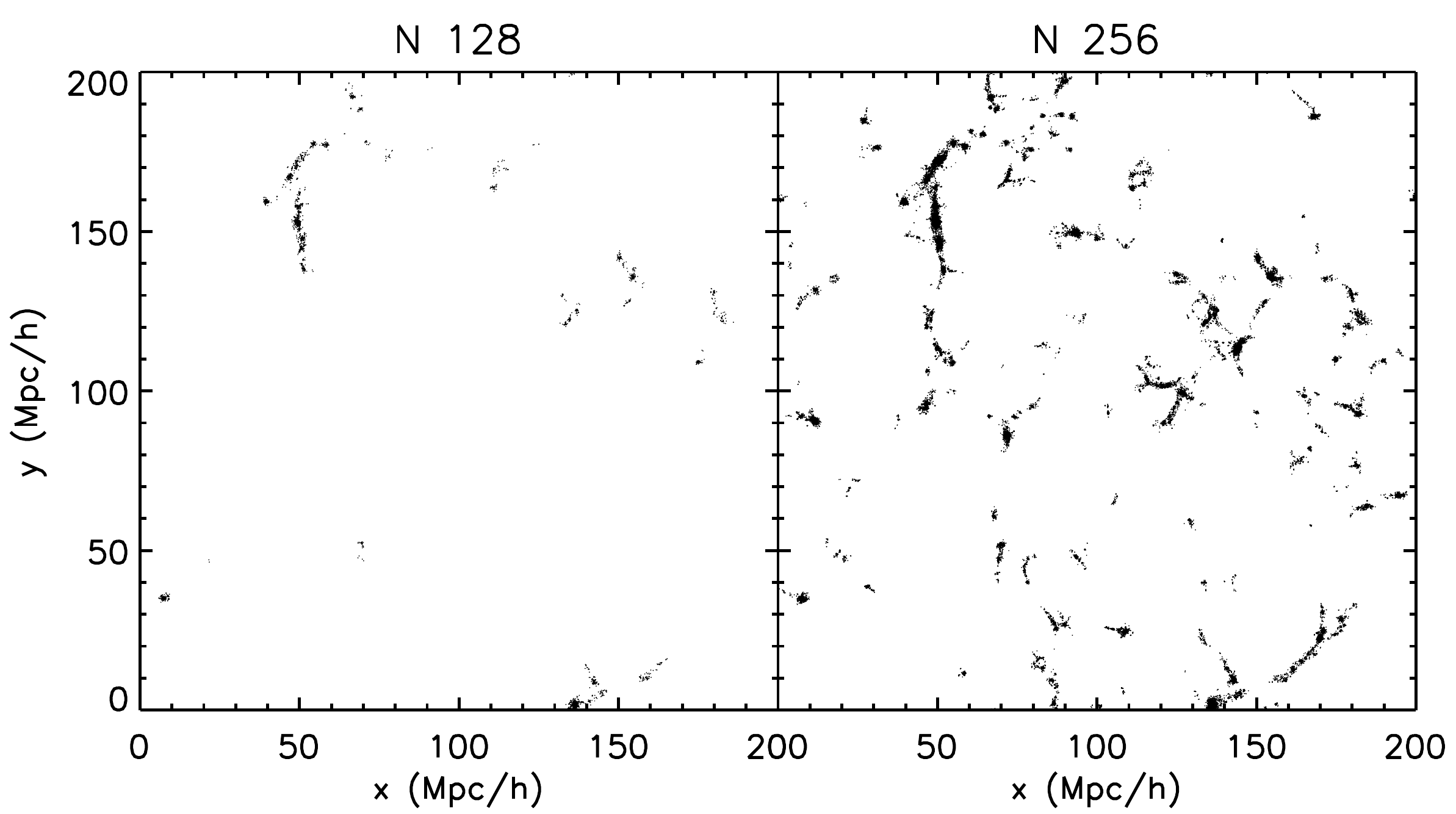}
\caption{A thin, 2\hmpc-thick slice through the 200\hmpc\ Planck simulations with $L/N=1.6$\hmpc\ (left) and 0.8\hmpc\ (right), showing only particles in the largest $M>1$ structure, which consists of halo and filament particles and spans the simulation in all three dimensions.
\label{fig:slice_struct2}}
\end{figure}

To make this clearer, similarly to Figure~\ref{fig:slice_vf1} we visualize the percolating $M>1$ structure in the two lowest resolution simulations, which are nearest to the percolation transition for the set of halo and filament particles, in Figure~\ref{fig:slice_struct2}. All halo and filament particles in the largest structure are plotted in a thin 2\hmpc-thick slice through the simulation. This clearly percolates (in this slice) in the higher resolution, $L/N = 0.8$\hmpc, simulation in the $x$-$y$ plane, shown in the right panel; though extent in $z$ is not shown, the structure fills the box in that dimension as well. The left panel, $L/N=1.6$\hmpc\ simulation is less clear from this slice, since it consists of much fewer particles covering a much smaller volume, but the structure does indeed occupy the entire simulation volume and is not localized to one area of the box. This structure nevertheless contains 40\% of all halo and filament particles, which themselves amount to 40\% of all particles at this resolution, so 16\% of all particles are in this structure.

\begin{figure}
\includegraphics[angle=-90,width=\hsize]{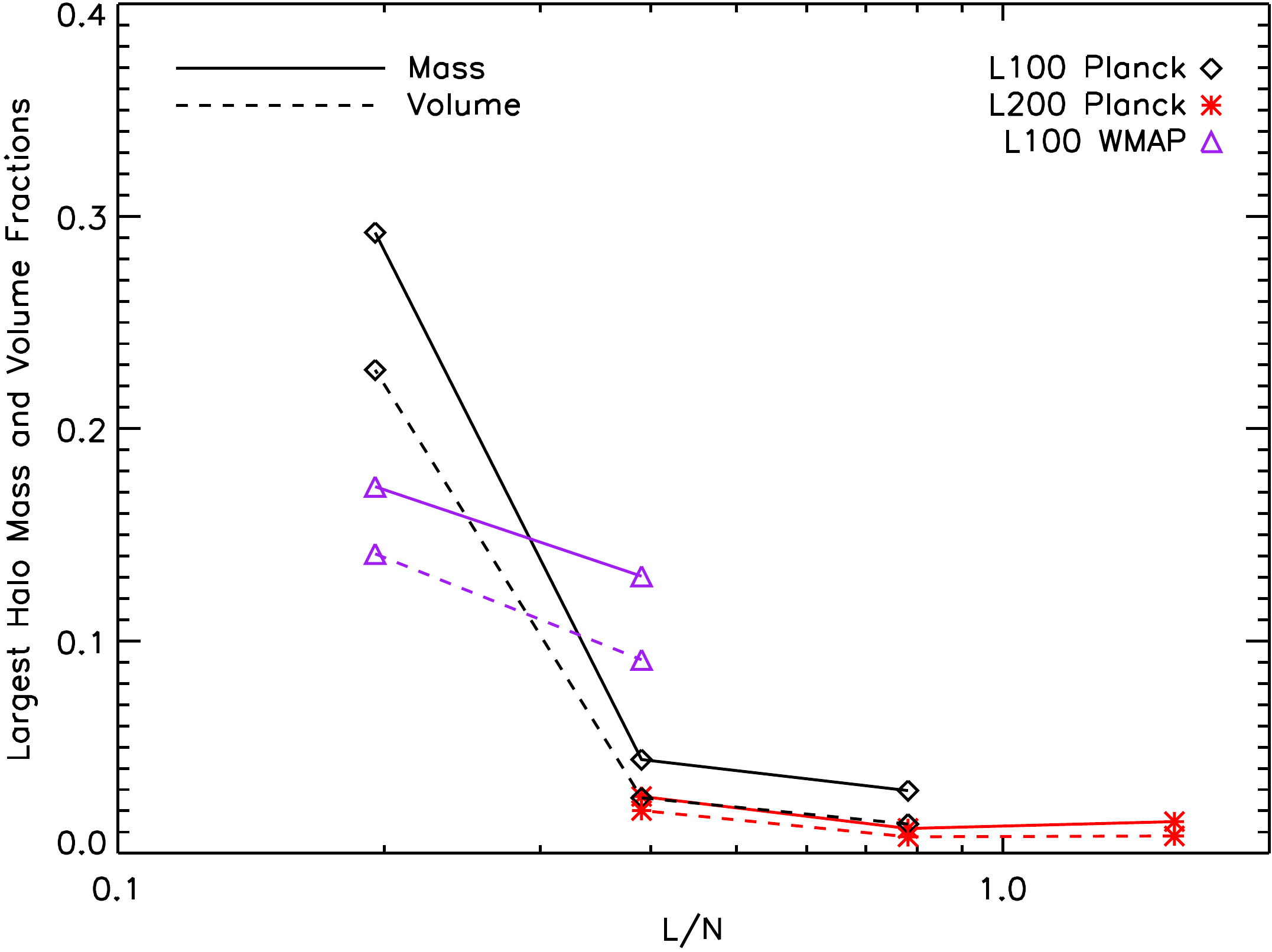}
\caption{Mass and volume fractions of the largest halo as a function of resolution. At the highest resolution, 57\% of the particles are in halos, and 30\% of these halo particles are in the largest connected structure, but this does not fully percolate the simulation volume in all three dimensions.
\label{fig:m3perc}}
\end{figure}

The network of halos in general does not percolate, but for the highest resolution simulations, the mass and volume fraction of the largest $M=3$ structure becomes quite large, as shown in Figure~\ref{fig:m3perc}. 30\% of the halo particles in the Planck $L/N=0.2$\hmpc\ simulation are in the largest halo structure, but this spans the volume in only two of the three axes and thus does not fully percolate. 
Note that because the extent of \org\ halos is defined by their outer caustic or first turn-around radius, they are much larger than friends-of-friends halos and can go out to 10 times the virial radius~\citep{Falck2012,Falck2014}.  These largest halo structures consist of nearby large groups connected by large filaments; as resolution increases, more halos in filaments become resolved, and \org\ halos can extend quite far so that they are tessellation neighbors of nearby halos.

\subsection{Shell-crossing Detection Efficiency}
\label{sec:shelleff}

A possible explanation for single-stream percolation -- i.e., that there aren't enough wall and filament particles to prevent void percolation -- could be that \org\ does not catch all the relevant shell-crossings. The two major ways this could happen are: (1), that particles cross at a higher redshift, then cross back during their collapse, to end at their initial configurations at $z=0$; and (2), that the four sets of axes used to detect particle reversals are insufficient. We investigate both below and find that they have a small effect, much smaller than the effect of ignoring the $f_v<1$ void boundaries as discussed in Section~\ref{sec:singleperc}, which does not prevent percolation. Thus, though \org\ misses a small fraction of shell-crossings, our conclusion of persistent percolation is unaffected.

We first determine how often particles cross at a higher redshift and return to their initial configuration at $z=0$, resulting in no crossing detected at $z=0$. We keep track of the morphology index, $M$, from 20 snapshots of a 200\hmpc, $256^3$ particle simulation ($L/N = 0.8$\hmpc), starting at $z=3.5$, and we do not allow a particle's morphology index $M$ to decrease in the next time step. This results in a change of mass fraction (i.e., fraction of particles with given $M$) of 0.017, -0.0058, -0.0027, and -0.0080 for halo, filament, wall, and void particles, respectively, by $z=0$. That is, there are slightly more halo particles and fewer wall, filament, and void particles by $z=0$ when taking into account each particle's $M$ since $z=3.5$. These very small changes in overall $M$ are unlikely to affect the percolation properties of the void particles, since this only decreases the fraction of void particles by 0.8\%, especially given that the set of $f_v=1$ void particles also percolates.

Secondly, in~\citet{Falck2012} we looked at the sufficiency of the four sets of axes used to detect shell-crossing by detecting shell-crossings on an additional six ``higher-order'' sets of axes (see Figure 2 of that paper for a schematic). We found that adding these axes resulted in only 5\% of particles increasing their morphology index, $M$, which includes void particles becoming wall particles, walls becoming filaments, and filaments becoming halos. For voids specifically, the fraction of void particles reduces by only 2\% (for the two WMAP simulation resolutions) when these extra axes are used to detect shell-crossings. Thus, looking for particle crossings along additional, rotated sets of axes only minimally increases the amount of shell-crossings detected, certainly not by enough to prevent the percolation of single-stream regions and having a negligible effect on the properties of the percolating void.

\subsection{The high resolution limit}
\label{sec:highres}

Since the fraction of collapsed mass, and particularly of halos, increases with resolution in a CDM cosmology, a natural question is that of what happens in the limit of infinite resolution. Does the fraction of single-stream voids become so low that percolation is prevented? What fraction of the mass is in halos or the multi-stream regime? Some clues can be gained by extrapolating the \org\ mass and volume fractions in Figures~\ref{fig:mfracs} and~\ref{fig:vfracs}. Grouping halos, filaments, and walls together as the collapsed mass, the void mass fraction is about 14\% at the highest resolution studied here, with an inter-particle separation $L/N=0.2$\hmpc, so the collapsed mass fraction is 86\%. Extrapolating the void mass fraction (lower right panel of Figure~\ref{fig:mfracs}) suggests a non-zero value at infinite resolution; however, this extrapolation does not quite agree with the finding of~\citet{Abel2012}, who measure 90\% of the mass in collapsed structures for a 40\hmpc\ $256^3$ particle simulation with $L/N=0.16$\hmpc. Linearly extrapolating between 14\% at $L/N=0.2$\hmpc\ and 10\% at $L/N=0.16$\hmpc\ gives a mass fraction of -6\% at infinite resolution, which is clearly unphysical. Note however that there is a slight disagreement between the WMAP and Planck void mass fractions at $L/N=0.2$\hmpc, and for small box sizes the results should be increasingly dependent on the random seed used to set the initial conditions.

The extrapolation of the mass fractions to infinite resolution is unclear, but the volume fractions tell a different story. In the bottom right panel of Figure~\ref{fig:vfracs} it is clear that there is only a small change of the void volume fraction as resolution increases, and the single-stream voids dominate the volume, perhaps approaching 80\% at infinite resolution. Since these volumes are defined from the Voronoi tessellation of dark matter particles, the mass fraction must be non-zero in order to have a non-zero volume fraction. The trend of the volume fraction suggests that single-stream percolation will persist even to infinitely high resolution; increasing resolution adds increasingly smaller structures, and these will not take up enough volume to separate single-stream regions into individual, distinct voids. Given the slow trend of void mass and volume fractions (and size of the largest void in Figure~\ref{fig:voidperc}) as $L/N \rightarrow 0$, very high resolution simulations are needed to answer this question with more certainty.

Theoretical models of structure formation either predict the fraction of collapsed mass in the infinite resolution limit or make some assumption about its value. The spherical collapse mass function of~\citet{Press1974} includes a normalization factor of 2 added by hand, which was later justified using excursion set theory and a sharp-$k$ filter~\citep{Peacock1990,Bond1991}. The inverse of this gives a prediction of 50\% of the mass in collapsed objects for a CDM power spectrum. In the ellipsoidal collapse model of~\citet{Sheth2001}, the normalization of their mass function was determined by assuming that this fraction is unity, or that all the mass is in bound objects of some mass~\citep[][p. 8]{Sheth2001}. \citet{Shen2006} combine triaxial collapse models with excursion set theory to predict the fraction of mass in the different cosmic web elements and find that for a $\Lambda$CDM cosmology, 99\% of the mass is in sheets of mass larger than $10^{10}M_{\astrosun}$ (which includes filaments and halos), 72\% in filaments (which includes halos), and 46\% in halos. Since this prediction is for a given mass and not down to the resolution limit, it seems likely that they also take the fraction of mass in objects of any mass to be unity.

A slightly different approach is taken by~\citet{Angulo2010}, who use excursion set theory to calculate the fraction of mass in objects down to the free-streaming scale of an 100 GeV neutralino. For a constant spherical collapse barrier, they predict 95\% of mass will be in a halo of any mass; for an ellipsoidal barrier, this reduces to 78\%. Though it is hard to extrapolate the halo mass fraction of Figure~\ref{fig:mfracs} to infinite resolution, it is definitely above 60\%, and a value of 80\% seems perfectly reasonable, in accord with the ellipsoidal collapse prediction of~\citet{Angulo2010}.

Theoretical predictions for the fraction of collapsed mass can be separated into the different cosmic web components using the Zel'dovich approximation. In this model, collapse occurs when at least one eigenvalue of the deformation tensor (the Hessian of the gravitational potential) is positive; for example, collapse along all three dimensions corresponds to all three positive eigenvalues, and collapse into a sheet corresponds to one positive eigenvalue. Setting $\lambda_1 > \lambda_2 > \lambda_3$, the probability that each eigenvalue is positive is
\begin{equation}
P(\lambda_1>0)=\frac{23}{25} , \; P(\lambda_2 > 0)=\frac{1}{2} , \; \& \; P(\lambda_3>0) = \frac{2}{25}
\end{equation}
\citep{Doroshkevich1970,Audit1997,Lee1998}. This corresponds to void, wall, filament, and halo mass fractions of 8\%, 42\%, 42\%, and 8\%, respectively. 
Clearly these halo, filament, and wall mass fractions do not agree with our $N$-body results in Figure~\ref{fig:mfracs}, but of course this is a rather simple linear approximation.


\section{ORIGAMI Voids}
\label{sec:voiddef}

The main result of this paper is that void boundaries are not well-defined by shell-crossed regions. Instead, a density-based criterion is needed to distinguish between separate regions of space that are expanding as voids; alternatively (or additionally), velocity information can also be used. Many void-finders start from underdense minima and grow voids until some maximum average density is reached, such as the canonical $\delta_V = -0.8$, or until the growing region runs into another growing void region, as in with watershed methods~\citep{Platen2007,Neyrinck2008}. Even with watershed void methods that do not rely on an explicit density parameter, some density threshold is often added to prevent the growth of voids into filaments and clusters. A density threshold greatly influences the resulting typical sizes of voids, and indeed if the parameter were set high enough (or not at all in the watershed case), the void regions (or the top-level watershed void) would percolate the volume~\citep{Neyrinck2008,Shandarin2006}.

In this section we describe an algorithm to identify a catalog of non-percolating \org\ voids. We describe our method of grouping \org-identified void particles by first identifying sets of void ``cores'' that are above a volume threshold parameter. We then discuss the combined effect that this threshold and the simulation resolution have on the properties of the resulting voids, focusing on which threshold and resolution combinations can lead to void percolation. Note that we do not identify the hierarchy of sub-voids within larger voids (just as the halo catalog does not pick out sub-halos~\citep{Falck2012}), though in principle this feature could be added to the algorithm. The important difference from other void finders is that these voids contain only \org-identified single-stream void particles; halo, filament, and wall particles are effectively ignored. However, we stress that this is not the only way of grouping together single-stream void particles; some other method could give more idealized voids.

\subsection{Method}

This algorithm operates much like an inverted version of the halo-grouping algorithm described in~\citet{Falck2012}. We use the Voronoi/Delaunay tessellation to obtain a density estimate and set of neighbors for each particle, as described above.
Adjacent void particles on the tessellation are grouped into the same void, but to prevent over-connected (i.e. percolating) void regions, we first connect only those particles that fall above a VTFE volume threshold, $V/\bar{V} = V_{th}$ (or equivalently, below a $\rho/\bar{\rho}$ threshold). These void ``cores'' form the seeds of the final voids, and void particles cannot belong to more than one void; thus the numbers and sizes of voids depend strongly on this volume threshold. However, it should not be confused with the fiducial void density $\delta = -0.8$, which many algorithms use to limit the growth of voids. The effect of varying this Voronoi density parameter is discussed in the next section.

Ideally, the Voronoi particle volume (or inverse density) threshold, $V_{th}$, would create a network of voids that does not percolate and that matches a visual inspection of the cosmic web. A volume threshold that is too low (density too high) will connect most of the void particles in one percolating void core, and all thresholds below this will also result in void percolation; alternatively, a volume threshold that is too high (density too low) will grow voids only around the rarest density minima.  The maximum volume threshold that creates a percolating void will thus depend on the cosmological parameters and possibly also on the random initial conditions of a given simulation, in addition to resolution. 

\begin{figure}
\centering
\includegraphics[angle=-90,width=\hsize]{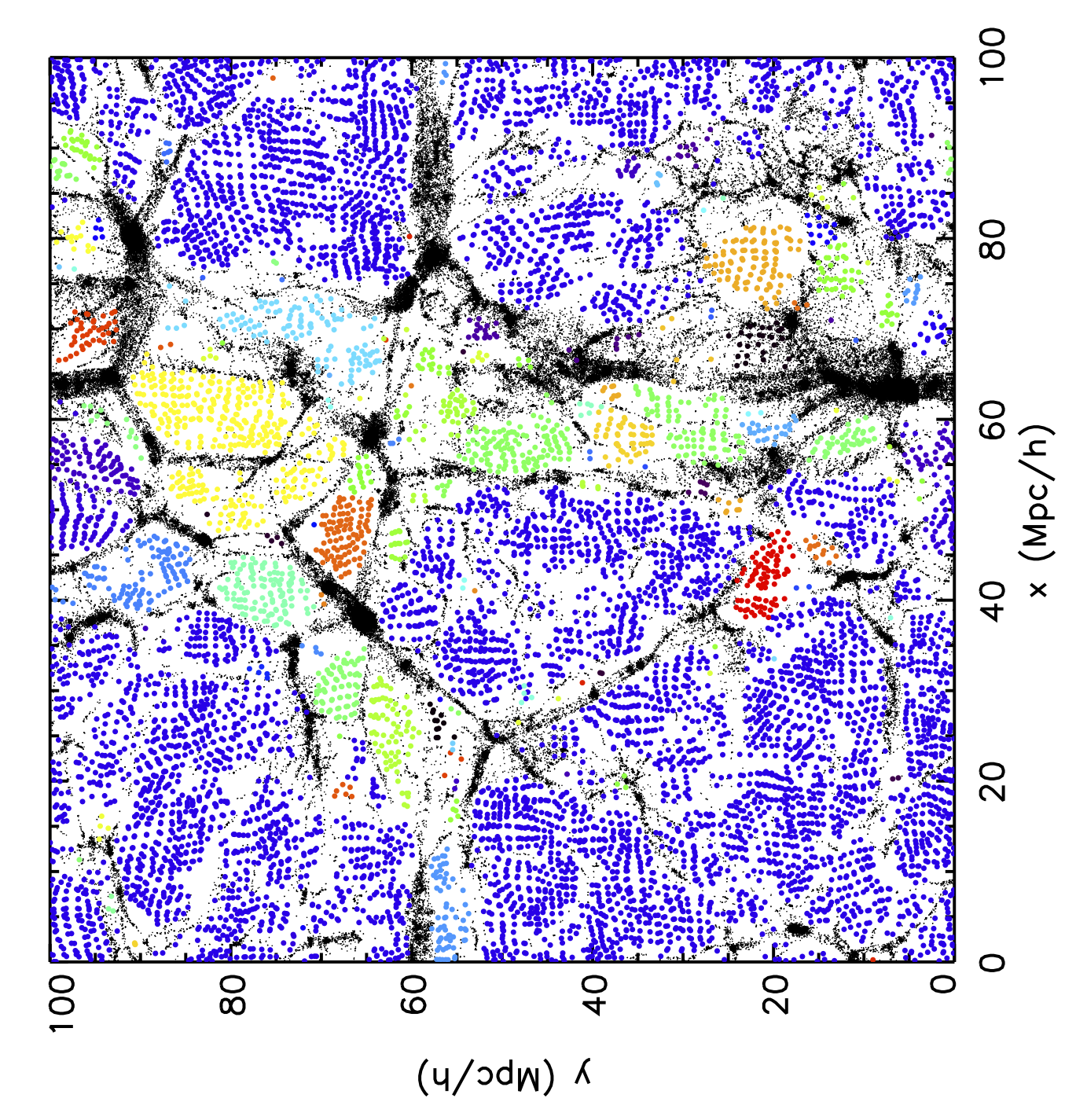}
\caption{The connectivity of void cores in a 2\hmpc\ thick slice through the WMAP $L/N = 0.4$\hmpc\ simulation for a Voronoi volume threshold of $V_{th} = 8$. The void particles for different void cores in the slice are given a random color and plotted as larger points over the wall, filament, and halo particles plotted as small black points. Most of the void particles in this slice are the same dark blue, corresponding to the percolating void ``core''.}
\label{fig:coreslice8}
\end{figure}

\begin{figure}
\centering
\includegraphics[angle=-90,width=\hsize]{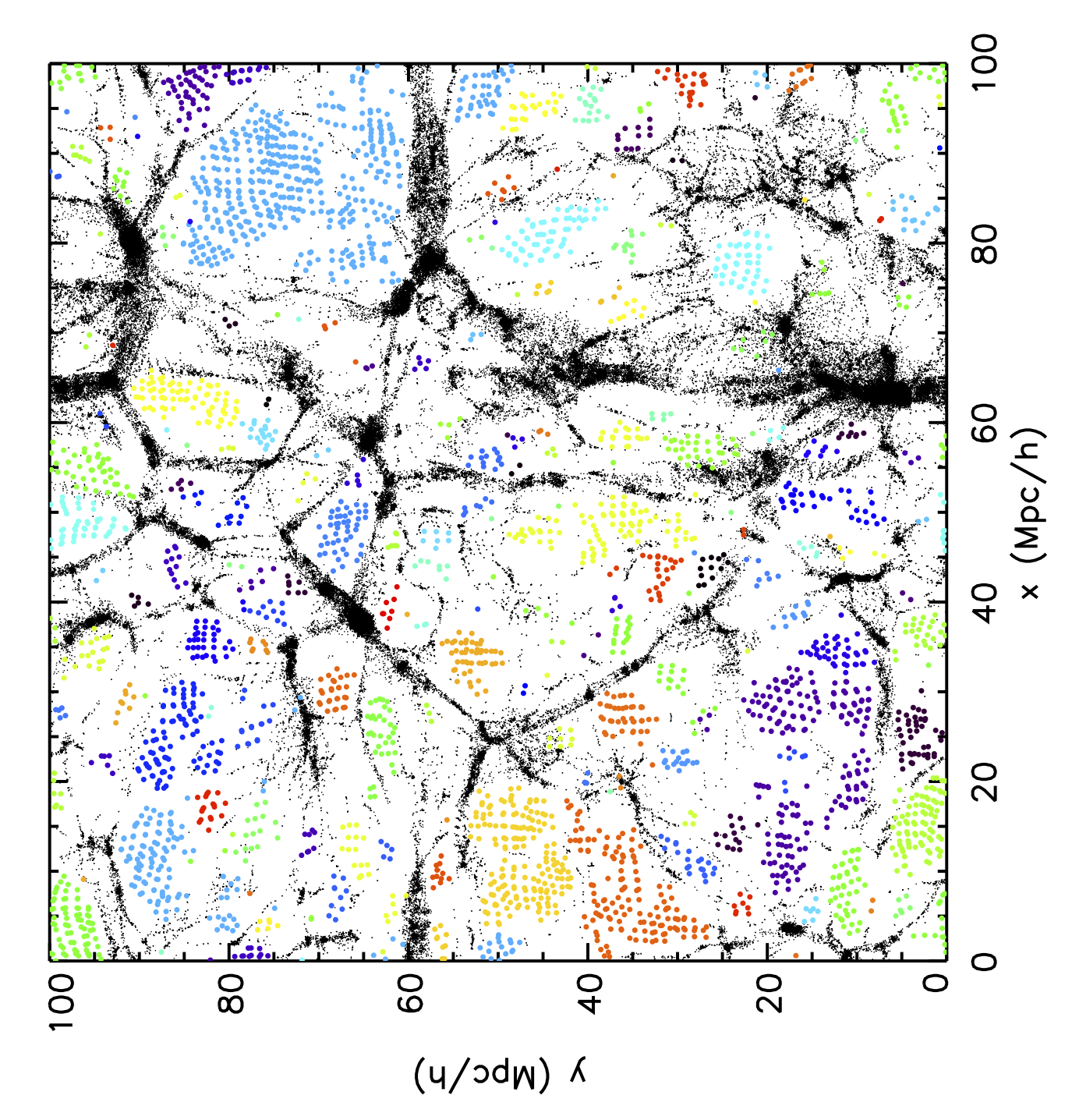}
\caption{The same as Figure~\ref{fig:coreslice8} but for a Voronoi particle volume threshold of $V_{th} = 12$. Void particles with volume above this threshold are plotted as large, randomly colored points over wall, filament, and halo particles in black. Percolation is prevented by defining void ``cores'' at or above this volume threshold.}
\label{fig:coreslice12}
\end{figure}

We show the connectivity of void particles above this Voronoi volume threshold, $V_{th}$, in Figures~\ref{fig:coreslice8} and~\ref{fig:coreslice12}. Both show 2\hmpc\ thick slices through the lower resolution ($L/N=0.4$\hmpc) WMAP simulation; wall, filament, and halo particles are plotted as small black dots, and void particles with $V/\bar{V}>V_{th}$ are plotted as larger points given a random color. Groups of nearby void particles with the same color are connected on the tessellation, i.e. they belong to the same void core. In Figure~\ref{fig:coreslice8}, $V_{th}=8$ (corresponding to $\delta = -0.875$), which is low enough that many void particles that satisfy this criterion are connected to each other on the tessellation and thus create a percolating void. In contrast, Figure~\ref{fig:coreslice12} shows void cores with $V_{th}=12$ (corresponding to $\delta = -0.92$), which are separated enough to prevent percolation (at this resolution). We will discuss in more detail the effect of simulation resolution and particle volume threshold on the connectivity and properties of the voids, not just the core $V/\bar{V}>V_{th}$ particles, in the next section.

\begin{figure}
\centering
\includegraphics[angle=-90,width=\hsize]{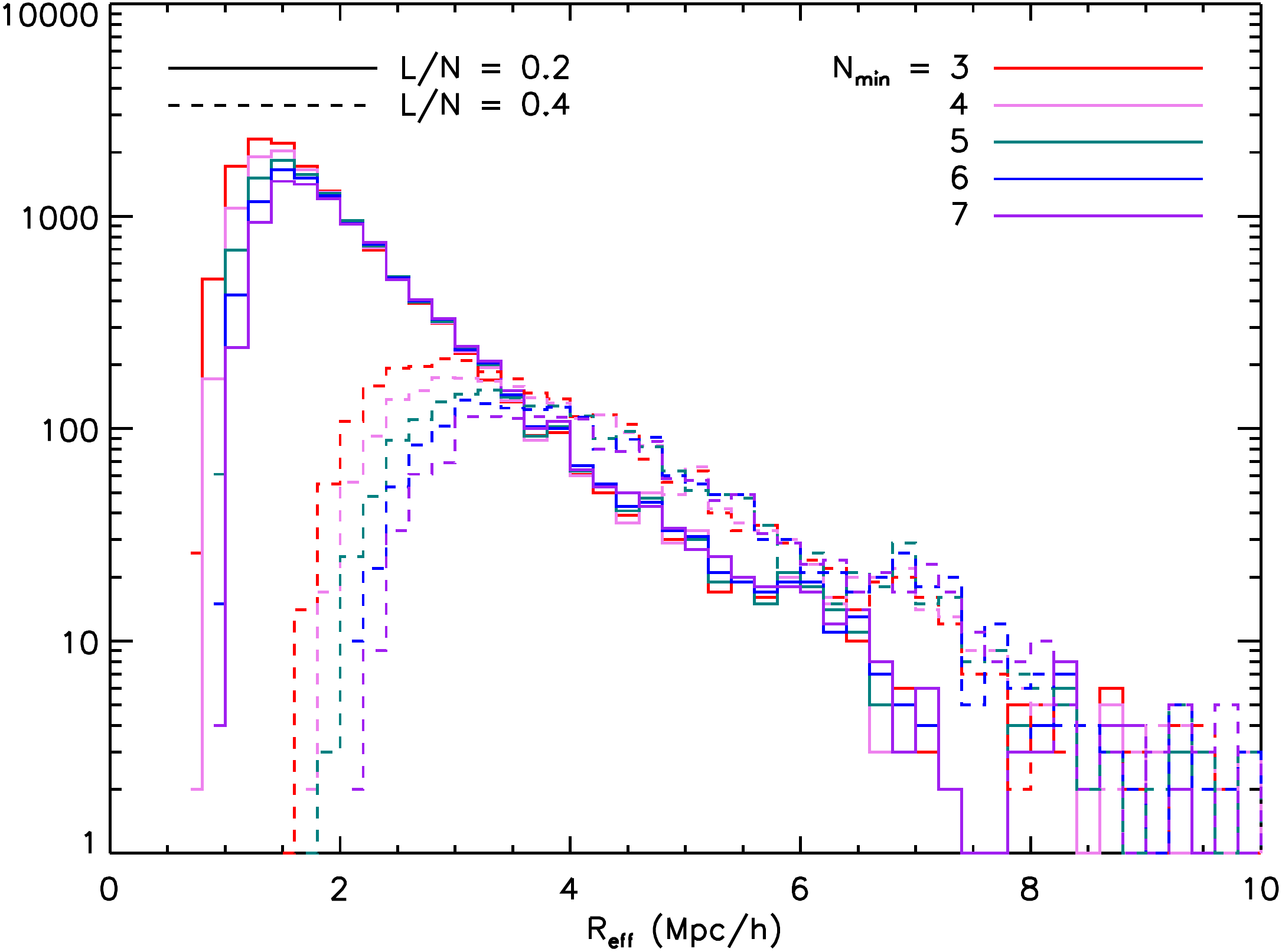}
\caption{Distributions of \reff\ for minimum core particle numbers of 3, 4, 5, 6, and 7, for both WMAP simulation resolutions and a Voronoi volume threshold of 15. Requiring a minimum number of ``core'' particles reduces the number of shallow density minima counted as voids, and the distribution of void sizes stabilizes at \reff$\gtrsim 10L/N$.}
\label{fig:reffncp}
\end{figure}

Once the initial set of void cores have been identified, the cores are pruned to prevent small, shallow volume peaks to be considered as real voids. We do this by removing cores with only a few particles, allowing the particles that make up these cores to be added to voids in the next step. The main effect of varying the minimum number of core particles is to decrease the number of very small voids as this minimum increases -- voids grown from these spurious cores are surrounded by void particles belonging to a real void and thus remain small. 
This is shown in Figure~\ref{fig:reffncp} for minimum core particle numbers of 3 to 7, for both WMAP simulation resolutions and the Voronoi volume parameter $V_{th}= 15$. We plot the distribution of void sizes defined as their effective radii, \reff$=(3V/4\pi)^{1/3}$, which is the radius of a sphere having the same volume as the (generally non-spherical) void. For the higher resolution, $L/N=0.2$\hmpc\ simulation, there is very little difference in the void radius distributions above \reff$\sim 2$ \hmpc; for the lower resolution, this increases to about 4 \hmpc. This suggests a rule of thumb that the smallest voids to trust in a given simulation should have \reff$\gtrsim 10L/N$. Note that we use a minimum core particle number of 5 as a default in what follows.

Once void cores have been found and spurious cores removed, we continue adding void particles that are neighbors on the tessellation to the void core particles, then neighbors of these neighbors, etc. until all connected void particles are added to voids. 
A small fraction (from 0.35\% at the lowest resolution to 2.0\% at the highest) of void particles end up not being added to any voids by this procedure. This is because either they are not adjacent to any other void particles on the tessellation or because they are not neighbors-of-neighbors of void core particles. These ungrouped void particles are thus in higher-density regions or surrounded by multi-stream particles.

Again we stress that these voids contain only single-stream, $M=0$ particles defined by \org. This means that they can surround small halos and filaments. There is also no density parameter that halts the growth of the voids; once the cores are identified, neighboring void particles are added iteratively until no more can be added. The volume of a void is defined as the total Voronoi volume of all its constituent particles; if they happen to completely surround any multi-stream particles, this is ignored.

\subsection{Void Properties}

\begin{figure}
\centering
\includegraphics[angle=-90,width=\hsize]{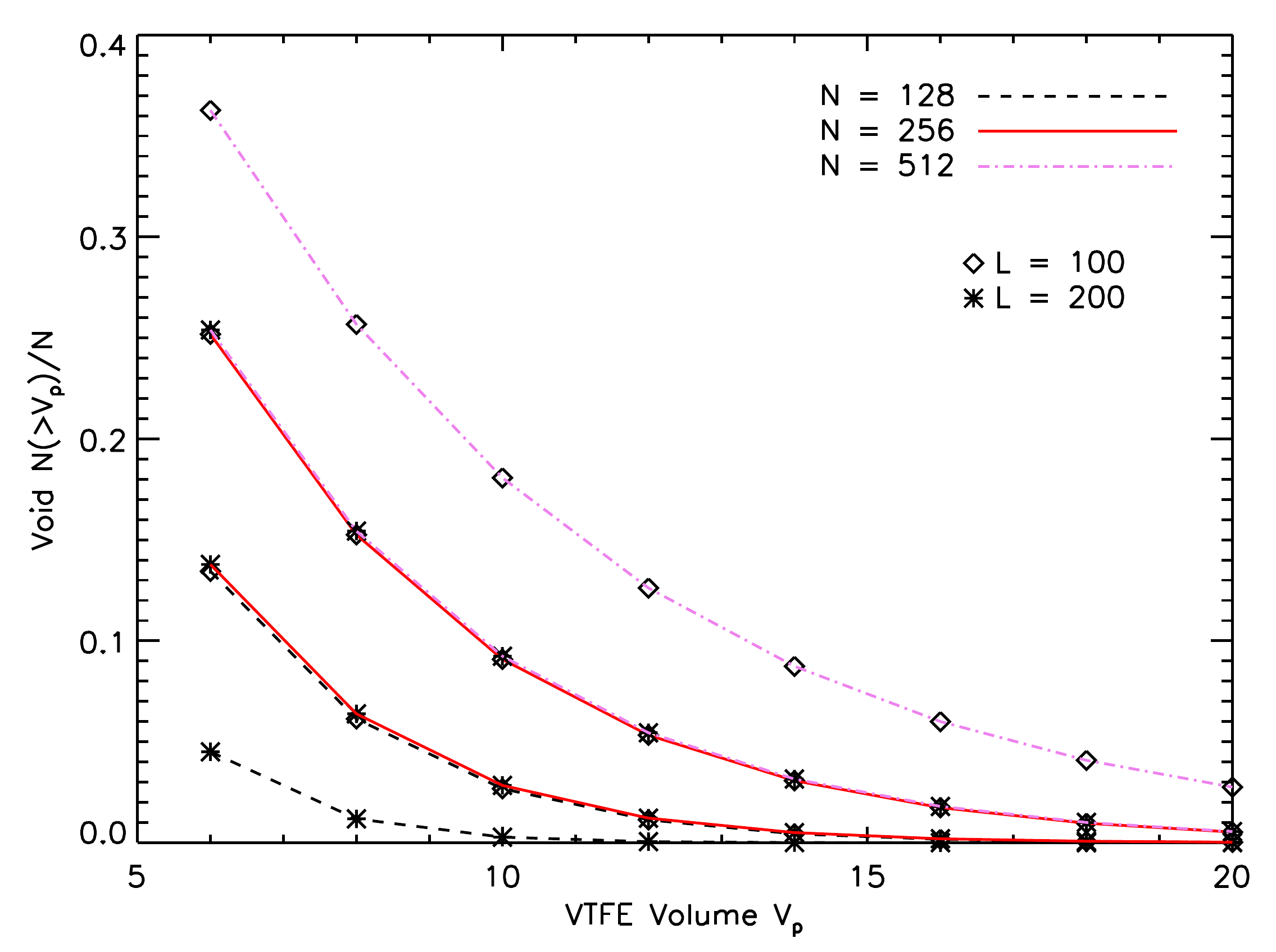}
\caption{The fraction of void particles having a VTFE volume above some threshold, for all Planck simulations.}
\label{fig:vcutnp}
\end{figure}

To understand the effects of the Voronoi particle volume threshold, $V_{th}$, and simulation resolution, $L/N$, on the void catalogs, we first show the fraction of void particles having a VTFE volume above the threshold, as a function of $V_{th}$, for each of the six Planck resolutions, in Figure~\ref{fig:vcutnp}. For a given volume threshold, the higher resolution simulations have a higher fraction of void particles with $V/\bar{V}>V_{th}$, even though they have a lower overall fraction of void particles (see Figure~\ref{fig:mfracs}). This is due in part to the scale-independent nature of the VTFE volumes -- as resolution increases, both the high density and low density regions are better resolved, resulting in a higher fraction of void particles having very low density. This is also reflected in Figure~\ref{fig:avgdens}, which shows how the average VTFE density of void, wall, and filament particles decreases as resolution increases, while the average density of halo particles increases.

Figure~\ref{fig:vcutnp} thus shows the fraction of void particles that make up void cores as a function of simulation resolution and volume threshold. This behavior drives much of the resolution and volume threshold dependence of the void catalogs. We apply the above method to create void catalogs for particle volume threshold values of $V_{th}=8$, 10, 12, 15, and 18.
For the lowest simulation resolution, the fraction of void particles with $V/\bar{V}>V_{th}$ is very low and indeed goes to zero for $V_{th} = 18$ (corresponding to $\delta=-0.94$). This means that no voids at all are found for that combination of resolution and volume threshold, and for this reason we exclude this $L/N=1.6$\hmpc\ simulation from the following analysis.

\begin{figure}
\centering
\includegraphics[angle=-90,width=\hsize]{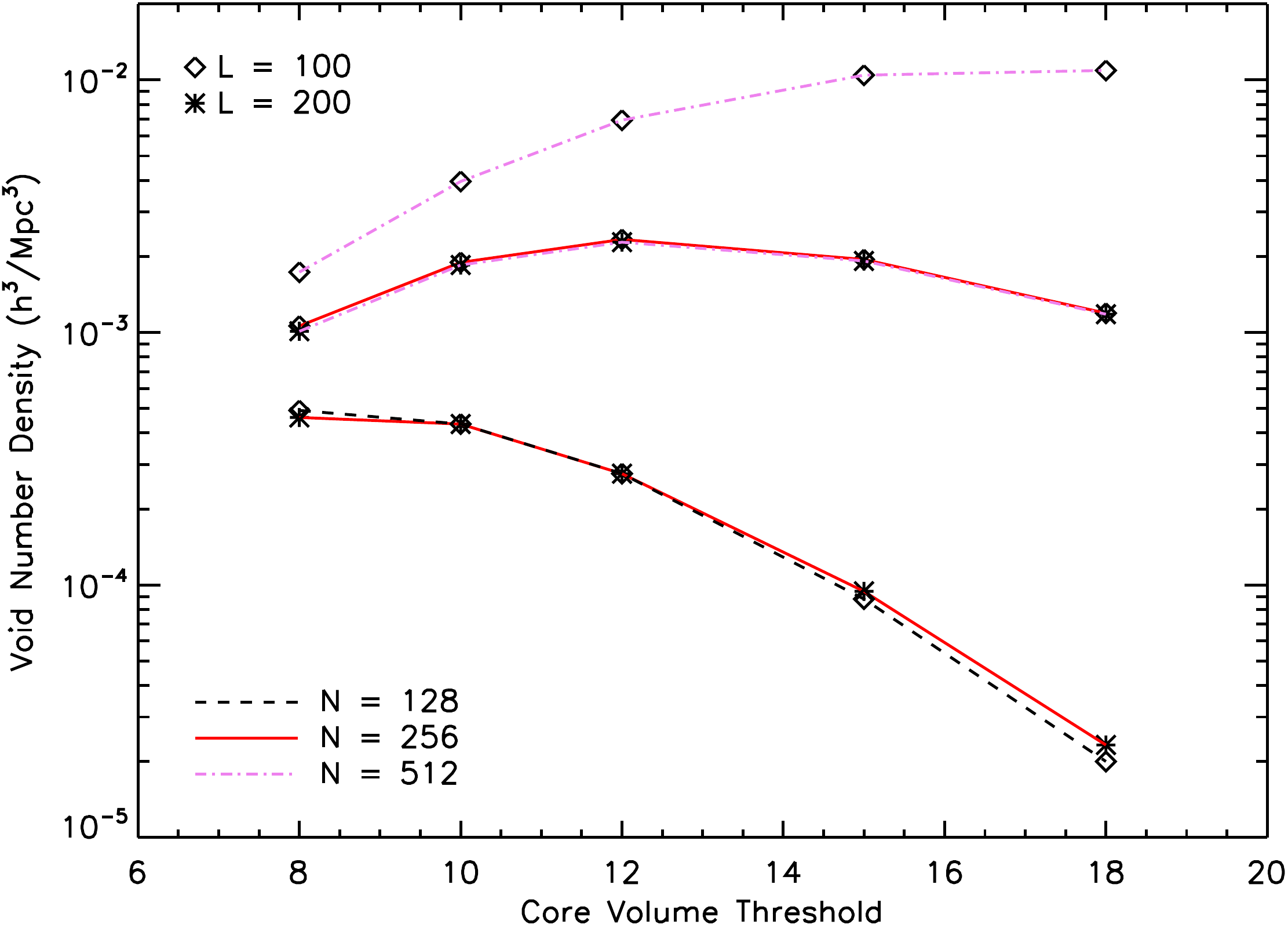}
\caption{The total number density of voids for five Planck simulations (excluding the lowest resolution) as a function of the Voronoi volume threshold, $V_{th}$. Single-stream percolation causes a lower void number density and depends on both volume threshold and simulation resolution.}
\label{fig:totnv}
\end{figure}

We show the total number density of voids as a function of volume threshold in Figure~\ref{fig:totnv}, for five of the Planck simulations. Since \org\ voids are formed by adding $M=0$ particles to low-density void cores, Figure~\ref{fig:totnv} equivalently shows the number density of void cores. As expected, for a given volume threshold, more voids are found in the higher resolution simulations, since more small voids are resolved (see Figure~\ref{fig:reffncp}). Fewer voids could signal the existence of a large or percolating void, especially keeping simulation resolution constant.

There are two transitions in the number density of voids as $V_{th}$ is varied, and these manifest differently according to simulation resolution. The first transition occurs at low $V_{th}$ and is evident in the highest resolution, $L/N=0.2$\hmpc, simulations (upper curves): percolation happens when $V_{th}$ is too low such that a high fraction of void particles have volumes above this threshold, resulting in a low-density ``core'' that itself spans the simulation volume and a low number density of voids. The second transition is evident in the lowest resolution, $L/N = 0.8$\hmpc, simulations (lower curves): there are fewer voids at a high $V_{th}$ because of the rarity of these particles (see Figure~\ref{fig:vcutnp}). There are simply fewer regions from which to grow the single-stream voids, and since nothing halts their growth, they can fill a large part of the simulation volume if there are sufficiently few cores. The middle resolution, $L/N=0.4$\hmpc, contains both of these transitions and has a peak number density of voids in the middle range of $V_{th}$. We can safely assume that if the range of volume thresholds were extended indefinitely, the number of voids in the high resolution simulations would drop as $V_{th}$ increases and the number of cores becomes too rare, and the number of voids in the lower resolution simulations would drop as $V_{th}$ decreases and the ``core'' begins itself to percolate.

\begin{figure}
\centering
\includegraphics[angle=-90,width=\hsize]{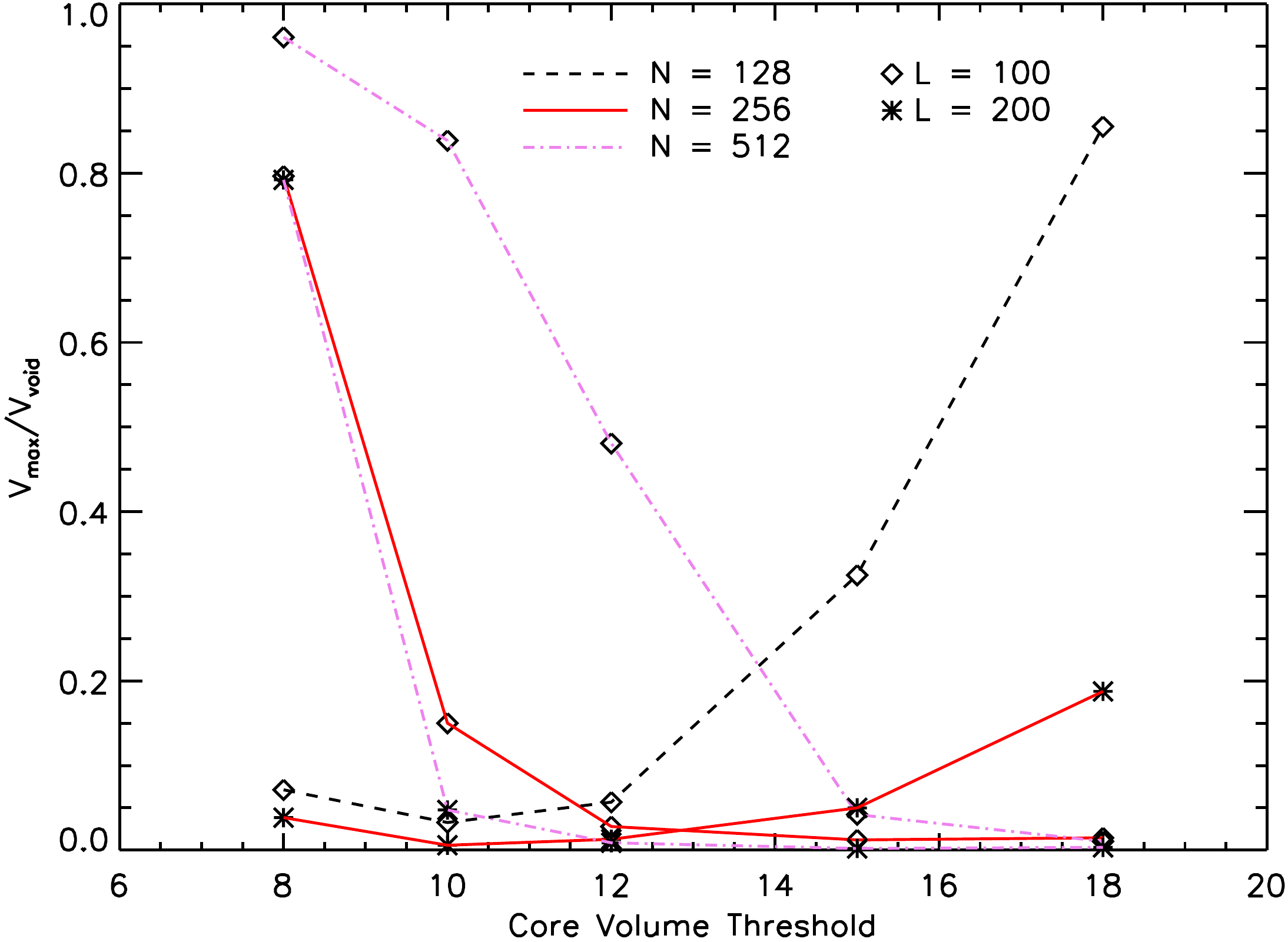}
\caption{Fraction of the total void volume in the largest void for five Planck simulations (excluding the lowest resolution) as a function of Voronoi volume threshold, $V_{th}$. Single-stream percolation occurs at high resolution and low $V_{th}$ (when the core itself percolates), and at low resolution and high $V_{th}$ (when high volume void centers are rare).}
\label{fig:maxvolfrac}
\end{figure}

To judge whether percolation occurs, as in Section~\ref{sec:percolation} we measure the fraction of the void volume in the largest void. This is plotted in Figure~\ref{fig:maxvolfrac} as a function of $V_{th}$ for five Planck simulations. 
Percolation clearly occurs for high resolution simulations with low $V_{th}$ voids, and for low resolution simulations with very high $V_{th}$, for the reasons explained above; there are also a few intermediate cases where it is unclear whether percolation has occurred from the volume fraction alone, though the volume fraction of the largest void is high. 

Though low resolutions and large volume thresholds can lead to voids that fill the volume due to the rarity of the $V/\bar{V}>V_{th}$ particles, this is more of an effect of the algorithmic definition (and indeed this can lead to zero voids found, as in the case of the $L/N=1.6$\hmpc\ simulation for $V_{th}=18$). The more interesting percolation is when the cores themselves span the volume, which happens at low values of $V_{th}$; for a given resolution, there is a transition from percolating to non-percolating as $V_{th}$ increases. The volume at which this transition occurs (though it appears gradual in Figure~\ref{fig:maxvolfrac}) increases as resolution increases, thus for a given $V_{th}$ percolation is more likely for a \textit{high} resolution simulation, and the volume fraction of the largest void is higher. Recall that when considering \textit{all} single-stream regions in Section~\ref{sec:percolation}, the volume fraction of the percolating void \textit{decreases} as resolution increases, which is the opposite of what we find here when restricting the density of the single-stream regions. In this case, void percolation is more likely at high resolution because more void particles have large volumes (e.g. Figure~\ref{fig:vcutnp}); this reflects the fact that voids found in the dark matter density have lower central densities than those found using halos or galaxies, which are sparser tracers of the density field~\citep[see, e.g.,][]{Colberg2008}.

As we found in Section~\ref{sec:multiperc} for the case of only halo particles, a large mass or volume fraction of the largest structure doesn't necessarily mean that structure percolates. We check whether the largest void spans the volume in all three $x$, $y$, and $z$ directions, and we find that this occurs for all combinations of resolution and $V_{th}$ where the volume fraction of the largest void is above 0.25 in Figure~\ref{fig:maxvolfrac}. For some combinations of resolution and $V_{th}$, the largest void spans the box in one or two, but not all three, dimensions.

\begin{figure}
\centering
\includegraphics[angle=-90,width=\hsize]{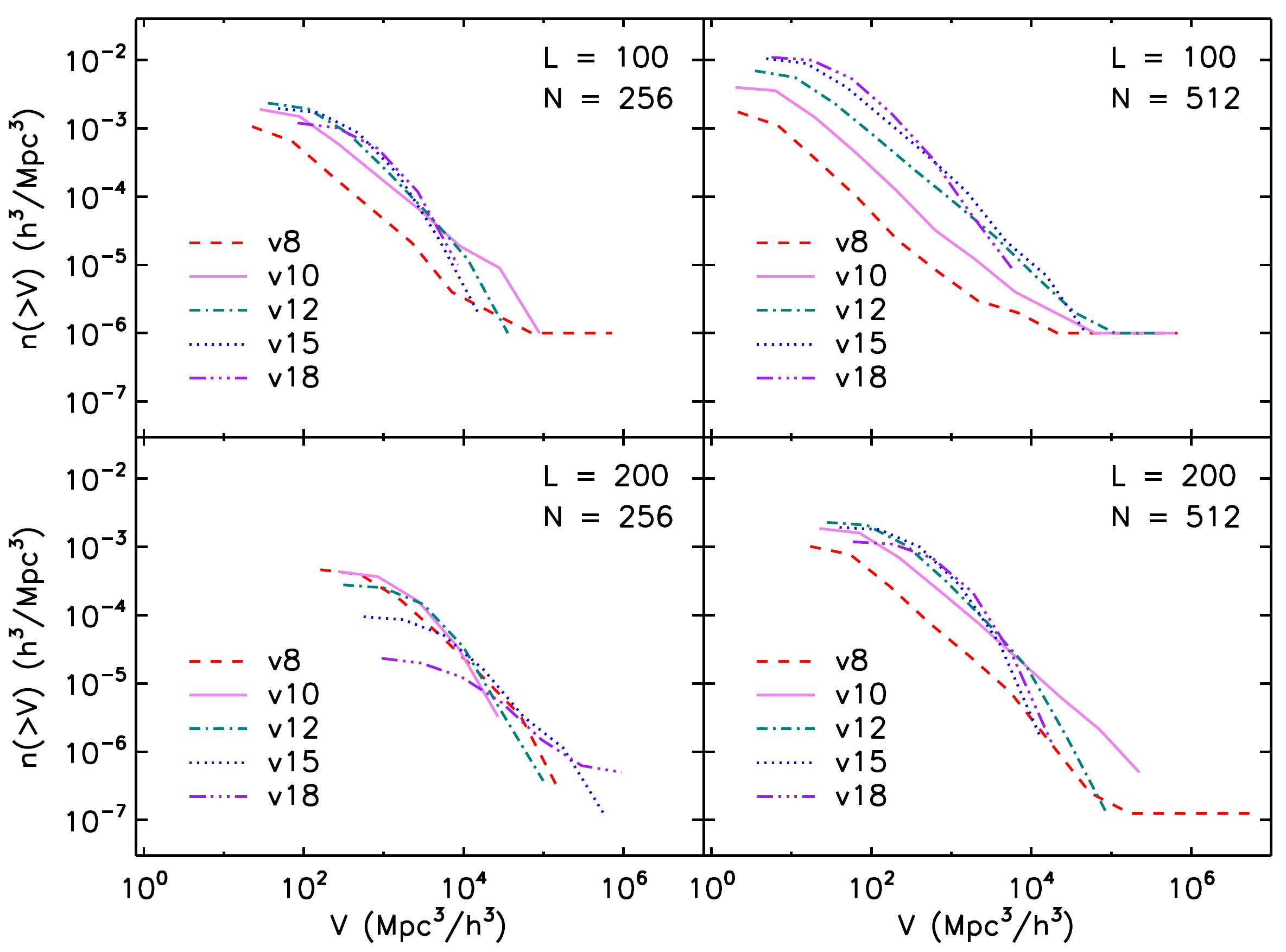}
\caption{Void cumulative volume functions for four Planck simulations and all Voronoi volume threshold values (denoted by different line styles and colors). Percolation causes the high volume end of this function to go flat, which never occurs for the $L/N=0.8$\hmpc\ simulation (lower left panel).}
\label{fig:vol}
\end{figure}

Similarly to the halo mass function, the void volume function characterizes the distribution of voids~\citep[e.g.,][]{ShethR2004,Jennings2013,Achitouv2013}. We plot the cumulative volume functions for all Voronoi particle volume thresholds and for the $N=256$ and $N=512$ versions of both Planck box sizes in Figure~\ref{fig:vol}. When percolation occurs, the largest void has a volume very near to that of the entire box, and the high end of the cumulative volume function goes flat. For the highest resolution simulation ($L/N=0.2$\hmpc, upper right panel), this occurs for $V_{th}$ values of 8, 10, and 12; for the two Planck simulations with $L/N=0.4$\hmpc, percolation occurs for $V_{th}=8$, and the shapes of the void volume functions look reasonably similar; while for the lower left panel with $L/N=0.8$\hmpc, percolation never occurs for the range of core volume thresholds tried. 

\begin{figure}
\centering
\includegraphics[angle=-90,width=\hsize]{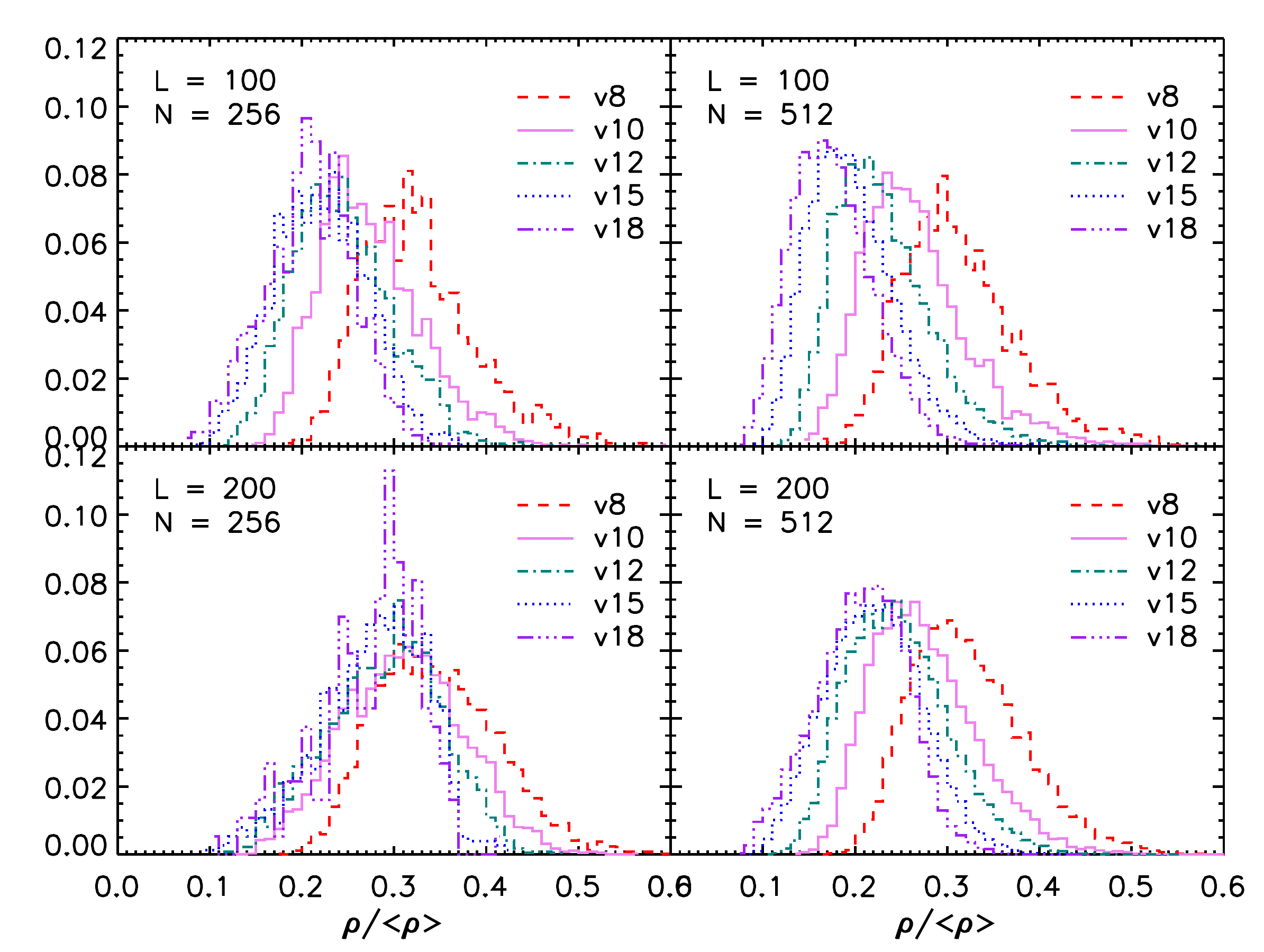}
\caption{The distribution of void densities $\rho/\bar{\rho}$ for four Planck simulations and all Voronoi volume thresholds (denoted by line styles and colors). Single-stream voids tend to have larger void densities than the canonical $\rho/\bar{\rho}=0.2$, even for non-percolating volume thresholds and resolutions, because they include the locally high density ridges that are nevertheless in the single-stream regime.}
\label{fig:vdens}
\end{figure}

Another interesting property to look at is the distribution of void densities. As opposed to the Voronoi volume threshold, which relates to the VTFE density of individual particles, the total void density is calculated here as the inverse of the average volume of all the particles in the void. Void density distributions are shown in Figure~\ref{fig:vdens} for all volume thresholds and for the $N=256$ and $N=512$ versions of both Planck box sizes (as in Figure~\ref{fig:vol}). It is notable that \org\ voids tend to have higher densities than the canonical $\rho/\bar{\rho} = 0.2$, which is often used to halt the growth of or to define voids. In contrast to these density-based void definitions, however, \org's dynamical definition of structures associates void regions with the absence of shell-crossing. As we have seen in~\citet{Falck2012} and in Section~\ref{sec:origami}, this leads to some void particles having rather high densities that overlap with the densities of some wall, filament, and even halo particles (Figure~\ref{fig:rhotag}). \org\ voids thus have boundaries that include regions of higher-density that are nevertheless in the single-stream region, thus \org\ voids have higher average densities than methods that halt void growth at $\rho/\bar{\rho} = 0.2$.

The general trend with volume threshold is for void densities to increase as the threshold decreases. When percolation does not occur, this is simply because the void density is a reflection of the core density. However, the volume thresholds that lead to percolation, in particular $V_{th}=8$ (red dashed in Figure~\ref{fig:vdens}), have much higher densities; this is because when voids percolate, the void particles that are left over tend to be in high-density regions, since they are separated on the tessellation from the percolating void. Note that in these distributions, the density of a percolating void is only counted once. 
As resolution increases, for a given $V_{th}$ (that does not produce percolation), there is not a large change in the distribution of void densities.

\subsection{Summary}

In this section, we described a method to produce catalogs of individual, non-percolating \org\ voids, which are composed exclusively of single-stream dark matter particles. This involves first identifying void ``cores'' as groups of particles with a VTFE volume above a volume threshold, $V_{th}$, then growing the void cores by iteratively adding void particles connected on the tessellation. In tandem with the simulation resolution, the volume threshold affects the number, sizes, and possible percolation of the resulting void catalog. Because the Voronoi tessellation provides a scale-free measure of the density field, increasing the simulation resolution results in a larger fraction of particles both with high and low densities; as seen in Figure~\ref{fig:avgdens}, the average density of halo particles increases, and that of void particles decreases, as resolution increases. For voids, this means that a higher fraction of void particles are above a given volume threshold for high resolution simulations (see Figure~\ref{fig:vcutnp}), thereby increasing the number of voids, lowering their average densities, and making percolation more likely than in lower resolution simulations.

\section{DISCUSSION}

We have studied the nature of single-stream voids and found that single-stream regions percolate, spanning the simulation volume in all three dimensions. This means that, contrary to the theoretical expectation~\citep[e.g.,][]{ShethR2004} and idealized cosmic web models~\citep{Icke1991,Neyrinck2014}, voids are not bounded on all sides by shell-crossing; rather, local under-densities are surrounded by locally higher-density ridges that have not yet entered the multi-stream regime to form walls or filaments, at least not on all sides of the under-dense void. We find that multi-stream regions (including all halo, filament, and wall particles) also percolate. The identification of halos, walls, filaments, and single-stream voids is performed by \org, which determines the number of orthogonal dimensions along which shell-crossing has occurred for each dark matter particle; halos have crossed along three, filaments two, walls one, and void particles have not undergone shell-crossing. 

\org\ relies on no smoothing scale, thus the identification of the cosmic web is limited only by the simulation resolution. In CDM cosmologies, this means that the fraction of collapsed mass, and in particular of halos, increases as the simulation resolution increases and smaller-scale structures are detected. In contrast, methods that identify the cosmic web via either the shear or tidal tensor have volume and mass fractions that depend strongly on smoothing scale and eigenvalue threshold for collapse~\citep[see, e.g.][]{Forero2009}. We find that the percolation of single-stream voids persists down to high resolution, at an inter-particle separation of $L/N=0.2$\hmpc. At this resolution, only about 14\% of the mass is in the single-stream regime, but this corresponds to about 82\% of the volume; while the single-stream mass fraction decreases with increasing resolution, it does not appear to approach zero, and the volume fraction decreases only slightly, approaching about 80\%, implying that single-stream percolation will persist even to the continuum limit of infinite resolution. By extrapolating void mass fractions to infinite resolution, \org\ provides an indication that not all mass is in collapsed structures even in the continuum limit. 

The percolation of single-stream regions indicates that walls, defined as having undergone one-dimensional shell-crossing, do not dominate the cosmic web at the current epoch and do not always correspond to the local density ridges that bound voids found in the density field. This is likely related to the suppression of growth in low-density areas, which become like a low-$\Omega_M$ universe; in a large-scale void, the evolution of sub-walls, filaments, etc. will be stunted compared to if they were in a mean-density environment, and indeed may evacuate as voids merge~\citep{Sutter2014}. Preliminary investigations indicate that the mass fraction of \org\ walls dominated those of halos and filaments at higher redshifts (but were dominated by voids) until the growth of halos took over at around $z=1$. The evolution of the \org\ cosmic web will be the subject of a future study (and see also~\citet{Cautun2014}).

Individual, isolated voids found in the density field are thus not bounded on all sides by multi-stream walls, filaments, and halos. To create a catalog of isolated single-stream voids, we introduce a volume threshold parameter which defines a set of void ``core'' particles according to their VTFE volume. Void particles above this threshold are first connected on the tessellation to define void cores, then void particles are added to the cores iteratively until all void particles belong to some void. If this volume threshold is too low, the set of void ``core'' particles will themselves percolate, similarly to how the set of over/underdense grid cells can create a percolating super-cluster/void network depending on the density threshold~\citep{Shandarin2004,Shandarin2006}. If the volume threshold is instead too high, there will only be a few void cores; these limits depend on the simulation resolution, which determines how well the density field is sampled in addition to the fraction of single-stream void particles identified. For combinations of volume threshold and resolution that do not result in percolation, we find that the average void densities are in general higher than the canonical $\rho/\bar{\rho}=0.2$ because they include locally high-density ridges which are still in the single-stream regime. 

\org\ relies on the information of the locations of dark matter particles in the Lagrangian grid, therefore it is not well-suited to studying the cosmic web in galaxy surveys where we lack that information. However, there are promising new methods being developed which allow a reconstruction of the initial conditions which are fully consistent with the observed galaxy distribution~\citep{Kitaura2013,Hess2013,Jasche2014}. Using such methods, it may be possible to determine the dynamical cosmic web morphology in which current galaxies reside by mapping out the shell-crossings in the reconstructed density field with \org.

\section*{Acknowledgements}

The authors are grateful to Gustavo Yepes and Miguel Aragon-Calvo for providing the suite of simulations used in this study. BF appreciates insightful comments from and discussions with Alexia Schulz, Raul Angulo, Frank van den Bosch, Tom Abel, and Stephane Colombi. BF acknowledges support from STFC grant ST/K00090/1. MN is grateful for support from a New Frontiers in Astronomy and Cosmology grant from the Sir John Templeton Foundation and a grant in Data-Intensive Science from the Gordon and Betty Moore and Alfred P. Sloan Foundations.  
Code and simulation data used for this paper can be obtained on request from the authors.

\bibliographystyle{hapj}
\bibliography{refs}

\label{lastpage}

\end{document}